\documentclass[journal=jacsat,manuscript=article]{achemso}
\SectionNumbersOn
\usepackage[version=3]{mhchem} % Formula subscripts using \ce{}

\usepackage{todonotes} % for comments
\usepackage{multirow}

\author{Changbin Im}
\affiliation[Ulm University]
{Institute of Electrochemistry, Ulm University, Albert-Einstein-Allee 47, 89081 Ulm, Germany}

\author{Björn Kirchhoff}
\affiliation[Ulm University]
{Institute of Electrochemistry, Ulm University, Albert-Einstein-Allee 47, 89081 Ulm, Germany}

\author{Igor Krivtsov}
\affiliation[Ulm University]
{Department of Organic and Inorganic Chemistry, University of Oviedo-CINN, 33006 Oviedo, Spain}
\affiliation[University of Oviedo-CINN]
{Institute of Electrochemistry, Ulm University, Albert-Einstein-Allee 47, 89081 Ulm, Germany}

\author{Dariusz Mitoraj}
\affiliation[Ulm University]
{Institute of Electrochemistry, Ulm University, Albert-Einstein-Allee 47, 89081 Ulm, Germany}

\author{Radim Beranek}
\affiliation[Ulm University]
{Institute of Electrochemistry, Ulm University, Albert-Einstein-Allee 47, 89081 Ulm, Germany}

\author{Timo Jacob}
\affiliation[Ulm University]
{Institute of Electrochemistry, Ulm University, Albert-Einstein-Allee 47, 89081 Ulm, Germany}
\alsoaffiliation[Helmholtz-Institute-Ulm]
{Helmholtz-Institute-Ulm (HIU), Helmholtzstr. 11, 89081, Ulm, Germany}
\alsoaffiliation[Karlsruhe Institute of Technology]
{Karlsruhe Institute of Technology (KIT), P.O. Box 3640, 76021, Karlsruhe, Germany}
\email{timo.jacob@uni-ulm.de}

\title{Structure and optical properties of polymeric carbon nitrides from atomistic simulations}

\abbreviations{IR,NMR,UV}

\begin{document}

\begin{tocentry}

Some journals require a graphical entry for the Table of Contents.
This should be laid out ``print ready'' so that the sizing of the
text is correct.

Inside the \texttt{tocentry} environment, the font used is Helvetica
8\,pt, as required by \emph{Journal of the American Chemical
Society}.

The surrounding frame is 9\,cm by 3.5\,cm, which is the maximum
permitted for  \emph{Journal of the American Chemical Society}
graphical table of content entries. The box will not resize if the
content is too big: instead it will overflow the edge of the box.

This box and the associated title will always be printed on a
separate page at the end of the document.

% TOC
%\begin{figure}[t]
%    \centering
%    \includegraphics[height=3cm]{figures/TOC.pdf}
%    \caption{Caption}
%    \label{TOC}
%\end{figure}

\end{tocentry}

\begin{abstract}
Detailed understanding of the structural and photophysical properties of polymeric carbon nitride (PCN) materials is of critical importance to derive future material optimization strategies towards more desirable optical properties and more photocatalytically active materials. However, the wide range of structural motifs found in synthesized PCNs complicates atomistic simulations that rely on well defined models. Performing hybrid DFT studies, we systematically investigate formation energy trends and optical properties of PCNs as a function of dimensionality, going from molecular oligomers over periodic sheet models to stacked crystals. Thermochemical calculations that take into account vibrational enthalpy and entropy contributions predict that a mixture of structural motifs including the melon string structure, poly(heptazine imide), and g-\ce{C3N4} motifs is stable under typical synthetic conditions. The degree of lateral condensation as well as stacking can reduce the bandgap while out-of-plane corrugation of the material increases both stability and the optical gap. The key result of this work is that already small domains of strongly condensed PCN are calculated to give rise to favorable optical properties. This result reconciles conflicting literature reports indicating that the thermodynamically favorable melon motif has a too large bandgap compared to experiments, while the g-\ce{C3N4} structure, for which bandgap calculations are in better agreement with experiments, does not agree with measured chemical compositions of PCNs. Finally, we postulate a new computational model for carbon nitride materials that encompasses the most important structural motifs and shows a bandgap of \textit{ca.}~2.9~eV.

%It is found that a similarity of thermodynamic stability between the competing structures indicates the mixture of structural motifs as a product during the most of synthetic conditions. 
%Three structural parameters are proposed to elucidate the relationship between the structure and the optical property.
%1) The degree of condensation is proportional to the chemical potential of ammonia that mainly contributes to the thermochemical stability and the optical property.
%2) The high stacking preference over the lateral growth highlights the finite-sized ribbon structure, which is supported by the result that the optical property of the infinite size rapidly converges within a few heptazine units.
%3) The size and the degree of condensation dependency of the corrugation play the pivotal role in both thermodynamic stability and the optical property.
%These results from the study suggest that the observed optical properties of the carbon nitride materials can be better explained by the finite size of the mixed structure composed of the strongly-condensed and the less-condensed domains that is thermodynamically more favorable than the infinite-sized and the individual model structure.  

\end{abstract}


\section{Introduction}

Polymeric carbon nitrides (PCNs) have recently gained considerable attention as a class of non-metallic photocatalytic materials.\cite{kessler2017functional} Among many potential applications, this new material class is currently being explored in particular for photo-induced hydrogen production,\cite{wang2009polymer,schwinghammer2013triazine,liu2014uniform,savateev2019ionic,zhang2019tailoring,teixeira2022overcoming} alcohol oxidation,\cite{ding2011synthesis,zhang2012photocatalytic,wang2021efficient,da2022selective} and the \ce{CO2} reduction reaction\cite{xia2019designing,xia2020improving} due to their intriguing photo-physical properties. PCNs feature optical gaps of \textit{ca.}~2.8 eV.

Typical PCN materials obtained by straight-forward thermal synthesis are semi-crystalline and contain a multitude of micro-structural motifs.\cite{lotsch2007unmasking,thomas2008graphitic,krivtsov2020water,mitoraj2021study,lau2022tour} This lack of crystallinity complicates structural characterization. However, if controlled reaction conditions can be enforced, certain motifs can become more prominent. For example, the "melon" structure, first described by Berzelius and Liebig\cite{https://doi.org/10.1002/jlac.18340100102,https://doi.org/10.1002/jlac.18350150306} consisting of strands of linearly polycondensed melamine monomers interconnected via hydrogen bonds, was successfully synthesized by Lotsch \textit{et al.} by thermal treatment of melamine in a closed reaction vessel at 630~$^{\circ}$C.\cite{lotsch2005thermal,lotsch2006triazines,lotsch2007unmasking} A mesoporous, non-hydrogenated graphitic carbon nitride (g-\ce{C3N4}) was postulated and explored for photocatalytic hydrogen production.\cite{goettmann2006chemical,wang2009metal} High condensation g-\ce{C3N4} model systems, where heptazine units are linked by tertiary amines into a 2D sheet structure, were suggested as theoretical models for PCNs by Lie and Cohen as early as 1989.\cite{liu1989prediction} Since then, calculations explored models based on g-\ce{C3N4} that exhibit desirable optical properties which are in good agreement with experimentally measured PCN samples.\cite{lau2022tour} However, the low hydrogen content of this structure is incompatible with elemental analysis of PCN samples, which are usually in better agreement with melon-type model systems. To this date, the idealized g-\ce{C3N4} structure has not been unambiguously identified in any synthesized samples. As a result, the exact structural characterization of PCNs is still an open question, and no single computational model on its own so far was able to fully explain the measured properties.

Combined theoretical and experimental efforts were able to shed light on some of the structural parameters of PCNs. Gracia \textit{et al.} for example found that corrugation can stabilize heptazine-based PCNs.\cite{gracia2009corrugated1} Li \textit{et al.} found that micro-structural features of PCNs such as H-bonding, interlayer spacing, and stacking configuration are susceptible to temperature variations during synthesis.\cite{li2018temp} Salt melts were explored for PCN synthesis and the templating effect from these salts was found to facilitate the synthesis of PCN structures with large cavities, such as poly(triazine imide) (PTI) and poly(heptazine imide) (PHI).\cite{bojdys2008ionothermal,algara2014triazine,lau2016rational}

%However, the fundamental understanding of those structures is still insufficient because of fractional formation mechanisms, limited structural knowledge in the crystalline domain, and the discorded optical gap between the theory and the experimental. Hence, a comprehensive research into the structure and the optical property is essential to reconcile both sides and to provide insight into the photophysics and the photochemistry of carbon nitride materials.

Uncertainty with respect to the structure of PCNs is one of several issues that theoreticians face when simulating this new material class. Further complications include the importance of accurately describing the optical properties, which necessitates the use of costly hybrid density functional theory (DFT) methods that augment regular generalized-gradient approximation (GGA) calculations with exact exchange contributions derived from Hartree-Fock wave function theory calculations on the Kohn-Sham orbitals.\cite{melissen2015relationship,steinmann2017challenges} Unfortunately, large system sizes are often required to model PCNs, which can make hybrid DFT calculations computationally unfeasible. Finally, PCNs are characterized by a multitude of intricate structural features such as stacking, edge boundaries, and corrugation which further increases the complexity of model systems.\cite{perdew1985density,melissen2015relationshipcorru3,melissen2016dft,steinmann2017challenges} The structural complexity of PCNs therefore hampers computational efforts aimed at improving our understanding of their fascinating properties, including strong excitonic effects,\cite{wei2013strong} charge accumulation,\cite{godin2017time,yang2019electron,li2021photodriven,adler2021photodoping,seo2021mechanism} and tuning of the optical properties towards more efficient visible light absorption.\cite{krivtsov2020water,kroger2021interfacial} 

Most notably, there is significant disagreement in literature between theoretical bandgap predictions and experimentally measured bandgaps. In particular, models for which bandgaps of 3~eV and lower are calculated typically feature a high degree of condensation while elemental analyses of synthesized materials with similar bandgaps show hydrogen contents that are incompatible with the theoretical models.\cite{ruan2021theoretical,dong2021bimetallic,cheng2021carbon,tang2022uncovering} The authors are convinced that this discrepancy is the result of the structural complexity of PCNs which is hard to represent properly in simulations.

In this study, we aim at postulating a more complete model system for PCNs. To this end, the relationship between structures and optical properties of PCNs are systemically investigated. Calculations initially focus on exploring individual parameters such as the degree of condensation, increase of dimensionality, and corrugation. To investigate these parameters and to model the polycondensation reaction from monomer to various PCN structures, the thermochemical approach introduced by Botari \textit{et al.}\cite{Botari2017} is adopted and extended by incorporating vibrational enthalpy and entropy contributions of the solid-state materials. By varying the chemical potential of \ce{NH3}, $\Delta\mu_{\ce{NH3}}$, refined thermochemical maps are obtained and reveal a thermodynamic route for the formation of the fully-deaminated PCN structure, typically referred to as g-\ce{C3N4}, via PHI as an intermediate.

The key finding of the present work is that already small domains of g-\ce{C3N4} embedded in less condensed PCN structures are sufficient to obtain bandgaps of 3~eV and lower. Thus, a comprehensive model system is postulated from a combination of the most important structural features explored in this work that affect the stability and optical properties of PCNs. This structure is thermodynamically feasible, shows a bandgap of \textit{ca.}~2.9~eV while retaining a low degree of condensation, and is predicted to form under experimentally accessible conditions. This model constitutes an important step towards a more holistic description of this fascinating material class in atomistic simulations.

\section{Methodology}
\subsection{Computational Details} \label{computational}
 All calculations were performed using the Vienna Ab initio Simulation Package (VASP) version of 5.4.4 which employs the projector-augmented wave (PAW) method.\cite{kresse1993ab,kresse1994ab,kresse1996efficiency,kresse1996efficient,kresse1999ultrasoft} A plane wave energy cut-off of 400 eV was used and the wavefunction was optimized to an accuracy of $10 ^{-6}$ eV. Atomic coordinates were relaxed until forces reached below $5 \times 10^{-2}$ eV/\r{A}. Gaussian-type finite-temperature smearing was employed with a smearing width of 0.01 eV. Long range interactions were treated with DFT-D3 dispersion correction.\cite{grimme2010consistent,grimme2011effect}
 
 The atomic and cell coordinates were relaxed using the exchange-correlation functional by Perdew, Burke, and Ernzerhof (PBE) within the generalized gradient approximation (GGA).\cite{perdew1996generalized} Accurate final energy results were then obtained by performing single-point calculations on the PBE-optimized structures using the hybrid functional by Heyd, Scuseria, and Ernzerhof with 25~\% exact exchange and a screening factor of 0.2 \r{A}$^{-1}$ (HSE06).\cite{krukau2006influence} The HSE06 functional was found to deliver a good description of the electronic properties of PCNs.\cite{melissen2015relationship,melissen2016dft,steinmann2017challenges}
 
Atomic coordinates of molecular models were optimized using the ORCA package version 4.21\cite{neese2012orca,neese2018software} using the PBE0 functional,\cite{perdew1996generalized,ernzerhof1999assessment,adamo1999toward} def2-TZVP basis set, def2/J auxiliary basis set, and DFT-D3 dispersion correction. Then, single point calculations were performed in VASP using the HSE06 functional to obtain final energy results. The pre-optimization in ORCA was found to be necessary because geometry relaxations in VASP regularly did not find the energetic minimum structures for molecules. In particular, corrugation is more pronounced in the ORCA optimized structures, which in fact also give lower total energy results after subsequent HSE06 single-point calculations in VASP. We assume that ORCA, with which we used finite basis sets and a hybrid functional for the relaxation, enabled symmetry breaking more easily compared to the GGA-based VASP plane wave calculations. To avoid spurious interactions between molecules and in case of 2D surface calculations, a vacuum layer of ($\sim20$~\r{A}) was used and dipole correction was enabled in the relevant directions (molecules: all directions, surfaces: perpendicular to surface). The Monkhorst-Pack scheme was adopted to create \textit{k} point grids for Brillouin zone integration.\cite{monkhorst1976special} A single, $\Gamma$-centered \textit{k} point was used for molecular structures. The vibrational contribution was calculated from force constant matrices using the VASP-Phonopy interface.\cite{togo2010first,chaput2011phonon,togo2015first} A data set of atomic coordinate files of the investigated structures in the VASP-POSCAR file format has been made available.\cite{changbin_im_2022_6953396}
 
\subsection{Thermochemistry}
For thermochemical analysis, we consider a simple thermal synthesis which uses melamine as the monomer. The thermochemical stability for each structure is assessed by the formation free energy $\Delta G_{\text{F}}(T,p)$ using the following equation:
\begin{eqnarray}
\Delta G_{\text{F}}(T,\emph{p}) = G_{\text{structure}}(T,p) - (n\ G_{\text{melamine}}(T,p) - m\ \mu_{\ce{NH3}}(T,p)).
\end{eqnarray}
\nolinebreak

Temperature ($T$) and pressure ($p$) are taken into account for the solid-state structures. The free energy of any particular structure under investigation, $G_{\text{structure}}$, and of a melamine monomer, $G_\text{melamine}$, are calculated as follows:
\begin{eqnarray}
G = E^{\text{total}} + F^{\text{vib}} + pV  \label{equfree}    
\end{eqnarray}
where $E^{\text{total}}$ is total energy, $F^{\text{vib}}$ is vibrational free energy. The contribution of vibrational degree of freedom can be obtained from phonon density of states $\sigma(\omega)$.
\begin{eqnarray} \label{equvibsolid}
F^{\text{vib}}(T,\omega) &=& \int \,d\omega F^{\text{vib}}(T,\omega)\sigma(\omega) \nonumber \\
&=& \frac{1}{2} \hbar\omega + k_\mathrm{B} T \ln \bigl[1 -\exp\Big(\frac{-\hbar\omega}{k_\mathrm{B} T}\Big) \bigr]
\end{eqnarray}
\nolinebreak
The gas phase chemical potential of ammonia, $\mu_{\ce{NH3}}(T,\emph{p})$, which depends on temperature and partial pressure, is an important term to connect simulations with experimental conditions. We adopt the expression for $\Delta \mu_\text{\ce{NH3}}(T,p)$ from Botari \textit{et al}.\cite{Botari2017} This expression for the calculated $\mu_\text{\ce{NH3}}(T,\emph{p})$ is based on the polyatomic ideal gas approximation composed of translational, rotational, vibrational terms:\cite{rogal2006ab,dill2010molecular,reuter2016ab}
\nolinebreak
\begin{eqnarray} \label{equchempoNH3}
\mu(T,p) &=& E_{\ce{NH3}}^{\text{total}} + E^{\text{ZPE}}_{\ce{NH3}} + \Delta\mu(T,p) + pV \\ 
\Delta\mu(T,p) &=& \mu_{\text{trans}} + \mu_{\text{rot}} + \mu_{\text{vib}} \nonumber \\
&=& -{\frac{1}{2}}k_\text{B}T \Big\{ \ln \Big[ \Big(\frac{2\pi m}{h^{2}}\Big)^\frac{3}{2}\frac{(k_\text{B}T)^\frac{5}{2}}{p} \Big] +\ln\Big(\frac{k_\text{B}T}{{\sigma}^{\text{sym}}B_{0}}\Big)-\ln\Big[1-\exp\Big({-\frac{\hbar\omega}{k_\text{B}T}\Big)\Big]} \Big\}
\end{eqnarray}
where $k_\text{B}$ is the Boltzmann constant, $m$ is the mass of ammonia, $\sigma^{\text{sym}}$ is the symmetry number, $B_{0}$ is the rotational constant, $\omega$ is the fundamental modes of the particle, $\hbar$ is the reduced Planck constant.

To validate our approach, we successfully reproduced published formation free energy trends for PCNs, see Figure \ref{chempo_dNH3} and \ref{chempomap_dNH3}.\cite{Botari2017} The published trends, however, neglect the influence of vibrational contributions on the free energy of both oligomers and solids. Significant changes can be observed when vibrational contributions are included (see Figure \ref{example_phase_update}).

Finally, the degree of condensation (DoC) is used as a descriptor throughout this work. It is calculated as the percentage quotient between the number of condensation points found in a particular structure and the product of the number of melem units - which are stable intermediates formed from the precursor melamine - that the structure is made up of times the maximum number of condensations that each melem unit in the structure can support (6):
\begin{equation}
    \text{DoC} = \frac{\text{\# of condensations}}{N_\text{melem} \times 6} \times 100.
\end{equation}

\subsection{Model Systems}
\begin{figure}[htbp]
    \centering
    \includegraphics[width=\linewidth/3*2]{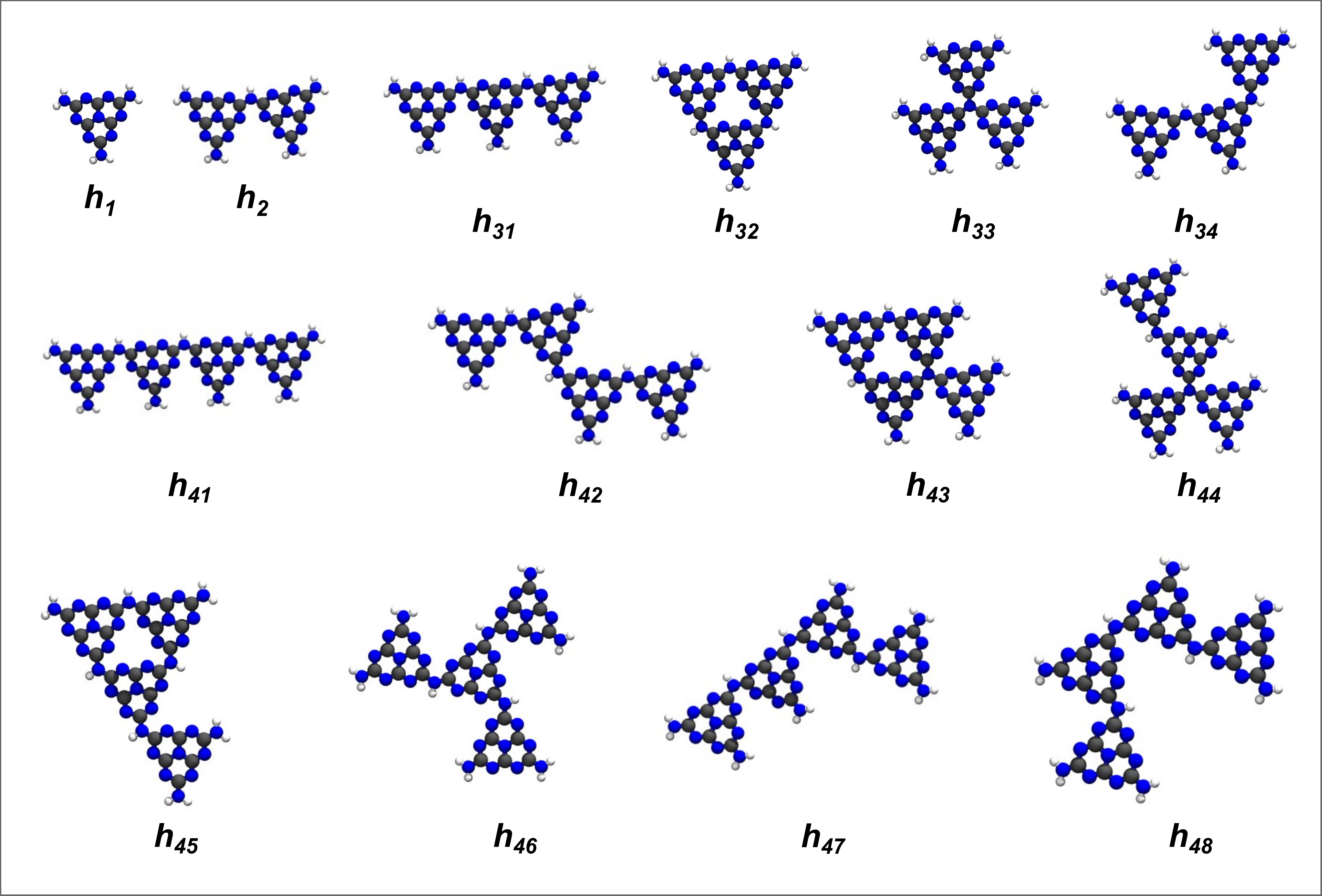}
    \caption{Illustration of molecular oligomer model systems investigated in this study. The models are named after a h$_{xy}$ naming scheme, where $x$ is the number of heptazine units and $y$ is an increasing number that indexes the oligomers.}
    \label{list_of_mol_structures}
\end{figure}

\begin{figure}[htbp]
    \centering
    \includegraphics[width=\linewidth]{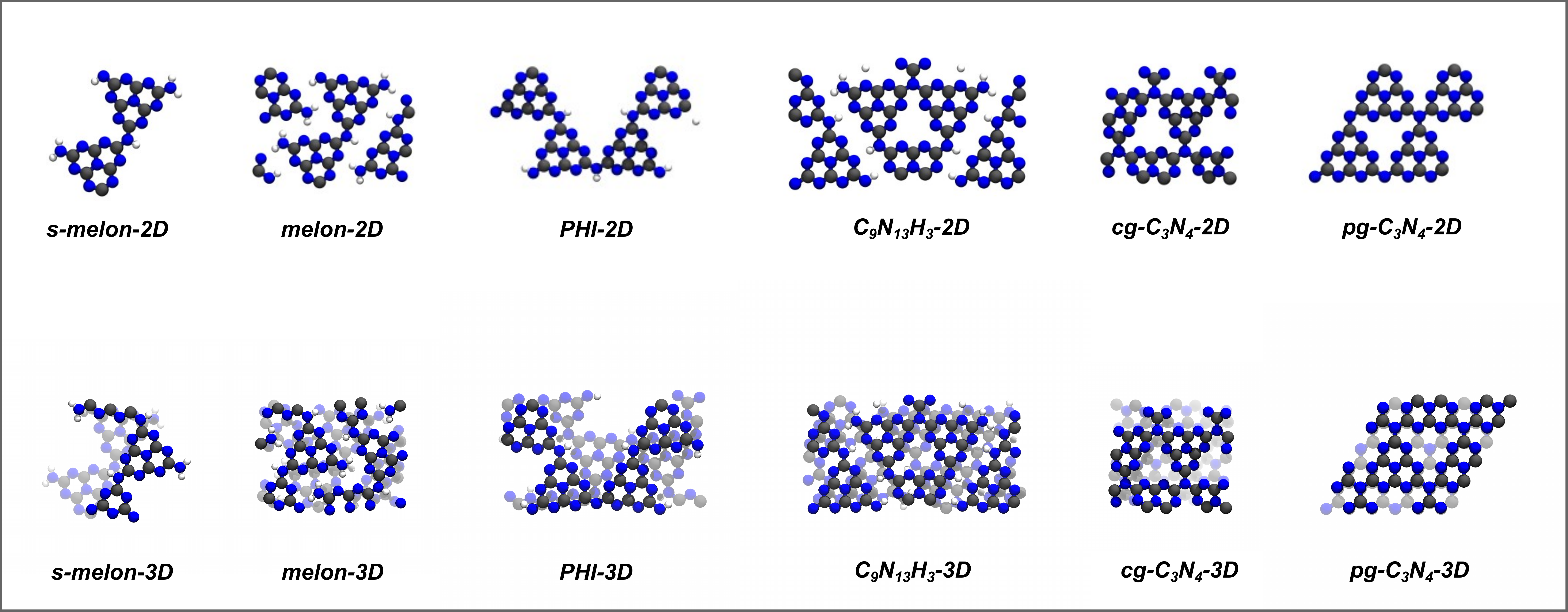}
    \caption{Illustration of periodic 2D surface and 3D crystal models investigated in this study. Structures are composed of s-melon, melon, PHI, \ce{C9N13H3}, cg-\ce{C3N4}(corrugated), and pg-\ce{C3N4}(planar) structural motifs.}
    \label{list_of_2d3d_structures}
\end{figure}

In order to simulate the thermodynamics of PCN formation, we simulate the polycondensation thermodynamics from the melamine monomer up to the tetramer phase. The investigated model systems are illustrated in Figure \ref{list_of_mol_structures}.
The investigated 2D- and 3D-periodic models are illustrated in Figure \ref{list_of_2d3d_structures}, while the mixed-motif structures postulated in section \ref{postulated} are shown in Figure \ref{list_of_mixed_structures}.
The atomic coordinates of all structures were relaxed using the methodology outlined in section \ref{computational}. Even though no constraints or restraints were placed on the models, the melon, pg-\ce{C3N4}, and \ce{C9N13H3} structures retained an idealized flat configuration, which is known to be a local minimum.\cite{gracia2009corrugated1,gao2020corrugation2,melissen2015relationshipcorru3} The other periodic structures show various degrees of corrugation. Corrugation is notorious due to the large number of local minima in the coordinate space. While this limitation affects our ability to predict absolute stability values with high confidence, it is only of minor importance when looking at larger scale stability trends, which is the focus of the present work.

\section{Results}

\subsection{Formation Energy Trends for 1D Materials}
\begin{figure}[htbp]
    \centering
    \includegraphics[width=\linewidth]{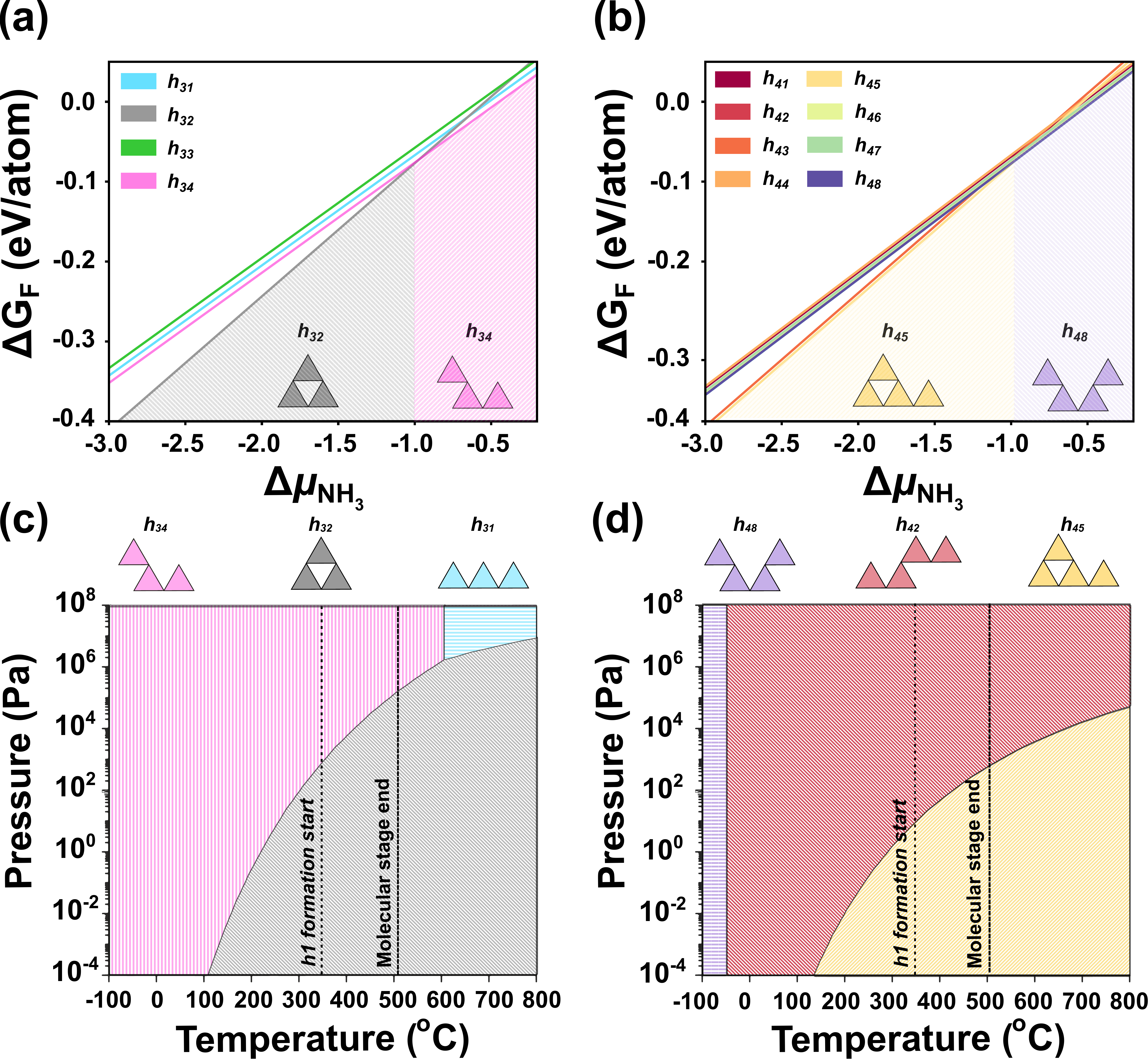}
    \caption{The free energy of formation for all (a) trimer and (b) tetramer structures are shown as function of $\Delta \mu_\text{\ce{NH3}}$. The free energy of formation is normalized to eV/atom to be able to compare structures with different total number of atoms. Free energy results are then expanded into $T$-$p_{\ce{NH3}}$-dependent phase diagrams for the most stable (c) trimer and (d) tetramer structures, respectively. The experimental temperature range for synthesis of these molecular structures is given between the dotted and the dashed line.\cite{li2018temp,yan2009temp2,lotsch2005thermal,lotsch2006triazines,lotsch2007new}}
    \label{phase_diagram_molecules}
\end{figure}

To understand the condensation process, formation free energy ($\Delta G_\text{F}$) phase diagrams are constructed using the melamine monomer and \ce{NH3} molecules as reference. Since there is only one possible configuration to form a dimer, different trimer and tetramer configurations are explored first. All calculated molecular structures are illustrated in Figure \ref{list_of_mol_structures}. The relative stability trends for the trimer and tetramer phases are shown in Figure \ref{phase_diagram_molecules}(a) and \ref{phase_diagram_molecules}(b), respectively.
The phase diagram reveals that two competing condensation processes exist during the molecular oligomerization stage: non-cyclic and cyclic condensation. Non-cyclic condensation is the dominant process at more positive values of $\Delta\mu_{\ce{NH3}}$ while cyclic condensation is dominant below $\Delta\mu_{\ce{NH3}}$ values of \textit{ca.} -1.0 eV, both in case of trimers and tetramers.

In case of the tetramer structures, the cyclic h$_{45}$ and the h$_{43}$ configurations (see Figure \ref{list_of_mol_structures}) are found to be similarly favorable. Both of these structures contain a condensed ring of three heptazine units (h$_{32}$) with an additional dangling heptazine unit, which is either attached at a terminal ($\ce{-NH}R$, where $R = \text{heptazine}, \text{h}_{45}$) or at a bridging group (\ce{-NR}, h$_{43}$). On the other hand, among the non-cyclic configurations, h$_{42}$ and h$_{48}$ compete as the most stable configurations with only small deviations of $\Delta G_\text{F}$ of $10^{-4} - 10^{-3}$ eV/atom. These structures as well as h$_{41}$ and h$_{42}$ actually constitute rotamers which can be converted into each other by rotation around bridging \ce{\textit{R}-NH-\textit{R}} moieties. Structure h$_{48}$ appears to be slightly more stable at $\Delta \mu_\text{\ce{NH3}}$ values close to zero, likely due to stabilizing van-der-Waals interactions that result from this configuration. Both of the angled structures (h$_{42}$ and h$_{48}$) are more stable than the linear structure h$_{41}$. However, the energy differences are small enough that a mixture of all of the discussed configurations may be formed from a thermodynamic point of view.

The $\Delta G_\text{F}$ differences between the most stable trimer and tetramer structures can be illustrated in a more practically-oriented way by expanding $\mu_\text{\ce{NH3}}$ into $T$-$p$ phase diagrams, see Figures \ref{phase_diagram_molecules}(c) and (d). At reported synthetic conditions,\cite{lotsch2005thermal,lotsch2006triazines,lotsch2007new} the formation of melem (h$_{1}$) begins at \textit{ca.}~350--370~$^{\circ}$C (Figure \ref{phase_diagram_molecules} (c) and (d)). From gravimetric analysis it is commonly assumed that formation of molecular oligomer structures is the dominant process until a temperature of \textit{ca.}~500~$^{\circ}$C.\cite{yan2009temp2} At higher temperature, more condensed structures (\textit{i.e.}, 2D-condensed polymers) are expected to form. From this analysis, it can be inferred that the more condensed structures h$_{32}$ (trimer) and h$_{45}$ (tetramer) are most likely to form in the typical experimental temperature regime (500--630 $^{\circ}$C)\cite{lotsch2007unmasking} and low \ce{NH3} partial pressure ($p_{\ce{NH3}}$), whereas the non-cyclic h$_{34}$ and h$_{42}$ structures are thermodynamically preferred at higher $p_{\ce{NH3}}$.

\subsection{Formation Energy Trends for 2D and 3D Materials}

\begin{figure}[htbp]
    \centering
    \includegraphics[width=\linewidth]{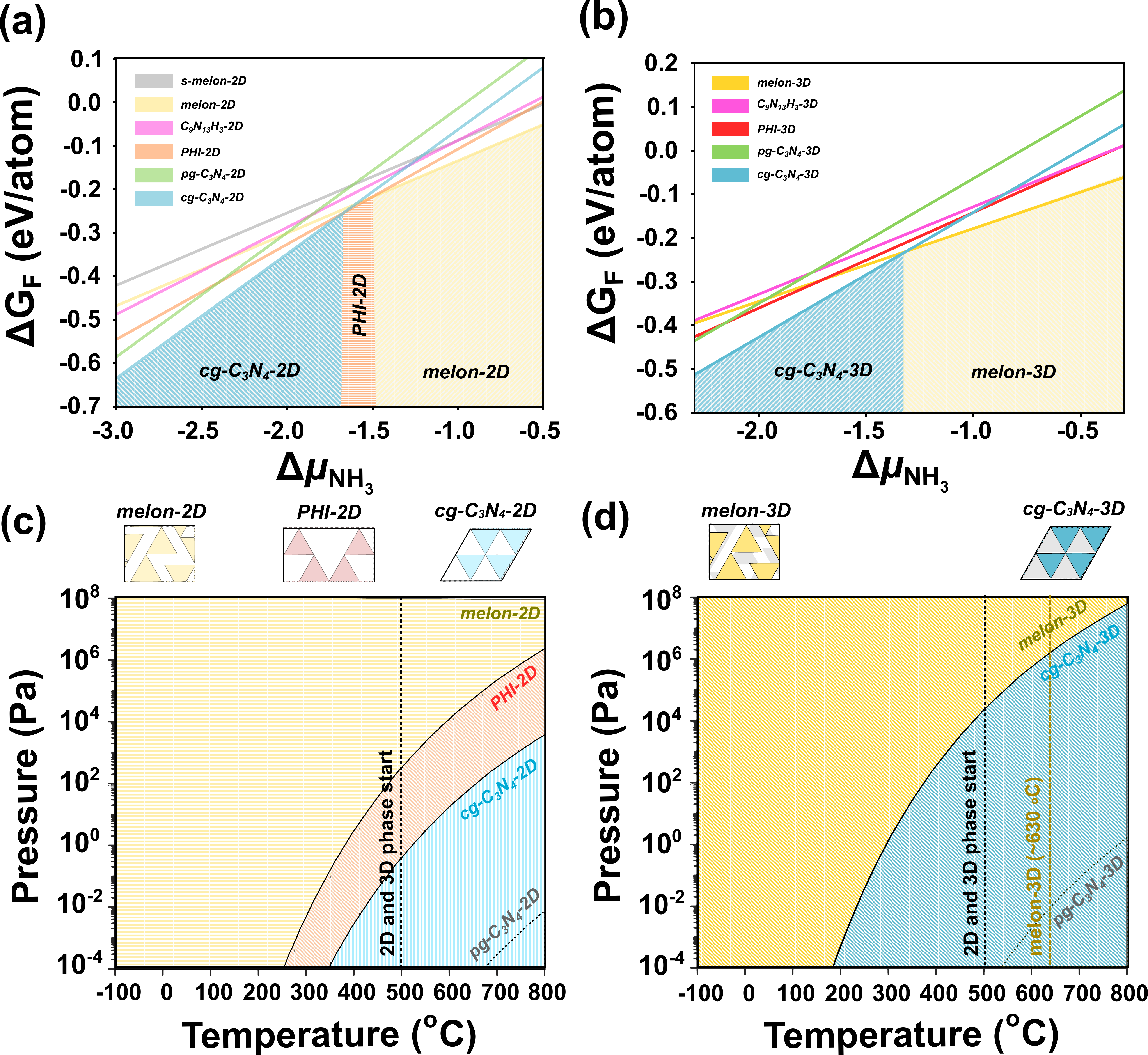}
    \caption{The free energy of formation for all (a) 2D and (b) 3D structures is shown as a function of $\Delta\mu_{\ce{NH3}}$. The free energy of formation is normalized to eV/atom to be able to compare structures with different total number of atoms. Then, free energy results are expanded into a $T$-$p_{\ce{NH3}}$-dependent phase diagrams for the most stable (c) 2D and (d) 3D structures, respectively. The experimentally measured temperature range for the formation and the decomposition of PCNs are given as dotted black and yellow lines.\cite{li2018temp,yan2009temp2} The dotted grey line indicates the stability region for the fully flat pg-\ce{C3N4}-2D/-3D phases which constitutes a local minimum structure and lays behind the more stable corrugated cg-\ce{C3N4}-2D/-3D structure.}
    \label{phase_diagram_2D3D}
\end{figure}

The formation free energy trends for 2D and 3D structures are shown in Figure \ref{phase_diagram_2D3D}.
The melon-2D structure accounts for the most stable structure up to $\Delta \mu_\text{\ce{NH3}} = -1.49$~eV. The PHI-2D structure appears as an intermediary phase between $\Delta \mu_\text{\ce{NH3}} = -1.49$ and $-1.68$~eV. For more negative $\Delta \mu_\text{\ce{NH3}}$, the highly-condensed cg-\ce{C3N4}-2D structure becomes the most stable phase. By deconvoluting $\Delta\mu_\text{\ce{\ce{NH3}}}$ into a $T$-$p$ diagram, Figure \ref{phase_diagram_2D3D} (c), a similar trend regarding the $p_{\ce{NH3}}$ is obtained as for the molecular phases, \textit{i.e.}, higher degree of condensation (DoC) can be achieved by reducing $p_\text{\ce{NH3}}$ at the experimental temperature conditions (500--630 $^{\circ}$C).\cite{lotsch2007unmasking}

In the 3D phase, cg-\ce{C3N4}-3D is the most stable structure at $\Delta\mu_\text{\ce{NH3}} \leq -1.33$~eV. At more positive values of $\Delta\mu_\text{\ce{NH3}}$, the melon-3D structure becomes more stable. The $T$-$p$ phase diagram (Figure \ref{phase_diagram_2D3D} (d)) reveals that there is thermodynamic competition between the formation of the cg-\ce{C3N4}-3D and melon-3D structures - or, in fact, their coexistence - in the experimental temperature range (500--630 $^{\circ}$C) and at a $p_{\ce{NH3}}$ of 1 atm. This phase border shows that lowering $p_{\ce{NH3}}$ will make the more condensed cg-\ce{C3N4}-3D structure thermodynamically more favourable.

\subsection{Optical Property Trends}

\begin{figure}[htbp]
    \centering
    \includegraphics[width=\linewidth*3/4]{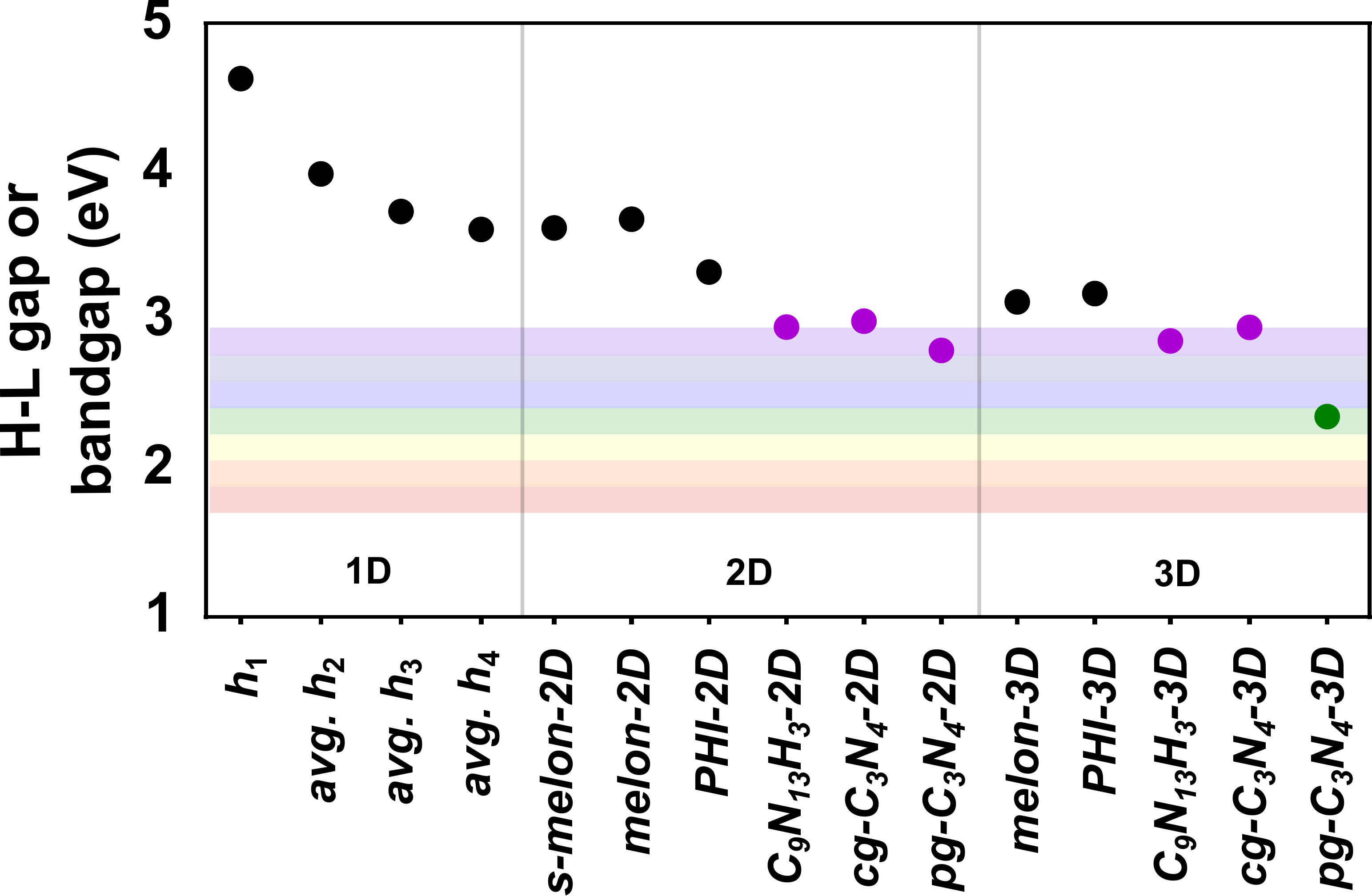}
    \caption{Summary of the calculated optical gap trends; HOMO-LUMO gaps are given for molecules and bandgaps are given for the periodic structures. For simplicity, the averaged HOMO-LUMO gaps are presented for the dimer up to tetramer structures. A color palate is given for reference to roughly indicate the energy range corresponding to visible light.}
    \label{bandgap_trend}
\end{figure}

%Figure \ref{phase_diagram_2D3D} established that the melon-3D structure, which consists of individual melon strings interconnected via hydrogen bonds, and fully condensed cg-\ce{C3N4}-3D structure are both likely to form under typical synthetic conditions (500--630 $^{\circ}$C, $p_\text{\ce{NH3}} \approx 10^5$~Pa). The question therefore arises which optical properties emerge from these different structural motifs. 
Calculated trends of the HOMO-LUMO gaps for the molecular structures as well as bandgap trends for the periodic structures are listed in Table \ref{table:table_optical_gap}. Figure \ref{bandgap_trend} summarizes the calculated values.
Note that HOMO-LUMO gaps for the molecular models are given as average values over all dimers, trimers, and tetramers, respectively. 

The optical property results give rise to several noteworthy trends: first, the optical gap decreases as the DoC increases. This trend is particularly apparent for the 2D periodic models. For instance, the melon-2D model, which shows a DoC of 50\%, has a \textit{ca.}~0.7~eV larger bandgap than the highly-condensed cg-\ce{C3N4}-2D model (DoC = 100\%). %Notably, some of the more strongly condensed tetramers exhibit a lower HOMO-LUMO gap (h$_{43}$: \textit{ca.}~3.5~eV) than some higher-dimensional structures such as the melon-2D structure (\textit{ca.}~3.7~eV). 

% I've deleted the comparison between HOMO-LUMO gaps and bandgaps because I'm afraid we might get into hell's kitchen for that. Just think about this: there can be materials with a bandgap of 0 eV but there can never be a molecule with a HOMO-LUMO gap of 0 eV. Both quantities say something about the optical properties but directly comparing them is super risky imo.
% I fully agree with you. CB

Second, as the dimensionality of the system increases from 1D molecules to 2D sheet models to 3D crystal models, the optical gap decreases. This trend is noteworthy in particular for the 2D to 3D transition where the growth mechanism is stacking of the 2D layers, \textit{i.e.} no additional covalent bonds are formed.

Third, corrugation of the structures increases the optical gap slightly. For example, the completely-flat pg-\ce{C3N4}-2D model, which constitutes a local minimum configuration of the system, has a \textit{ca.}~0.2~eV smaller bandgap than the corrugated and thermodynamically more stable ($\Delta E = 0.7$~eV/heptazine unit) cg-\ce{C3N4}-2D system. This effect is even more pronounced for the 3D crystal models, where the non-corrugated system has a narrower bandgap of \textit{ca.}~0.6~eV. At the same time, the interlayer spacing of the non-corrugated model decreases by \textit{ca.}~0.17~\AA\ compared to the corrugated model, indicating significantly stronger interaction between layers.

\begin{figure}[htbp]
    \centering
    \includegraphics[width=\linewidth/3*2]{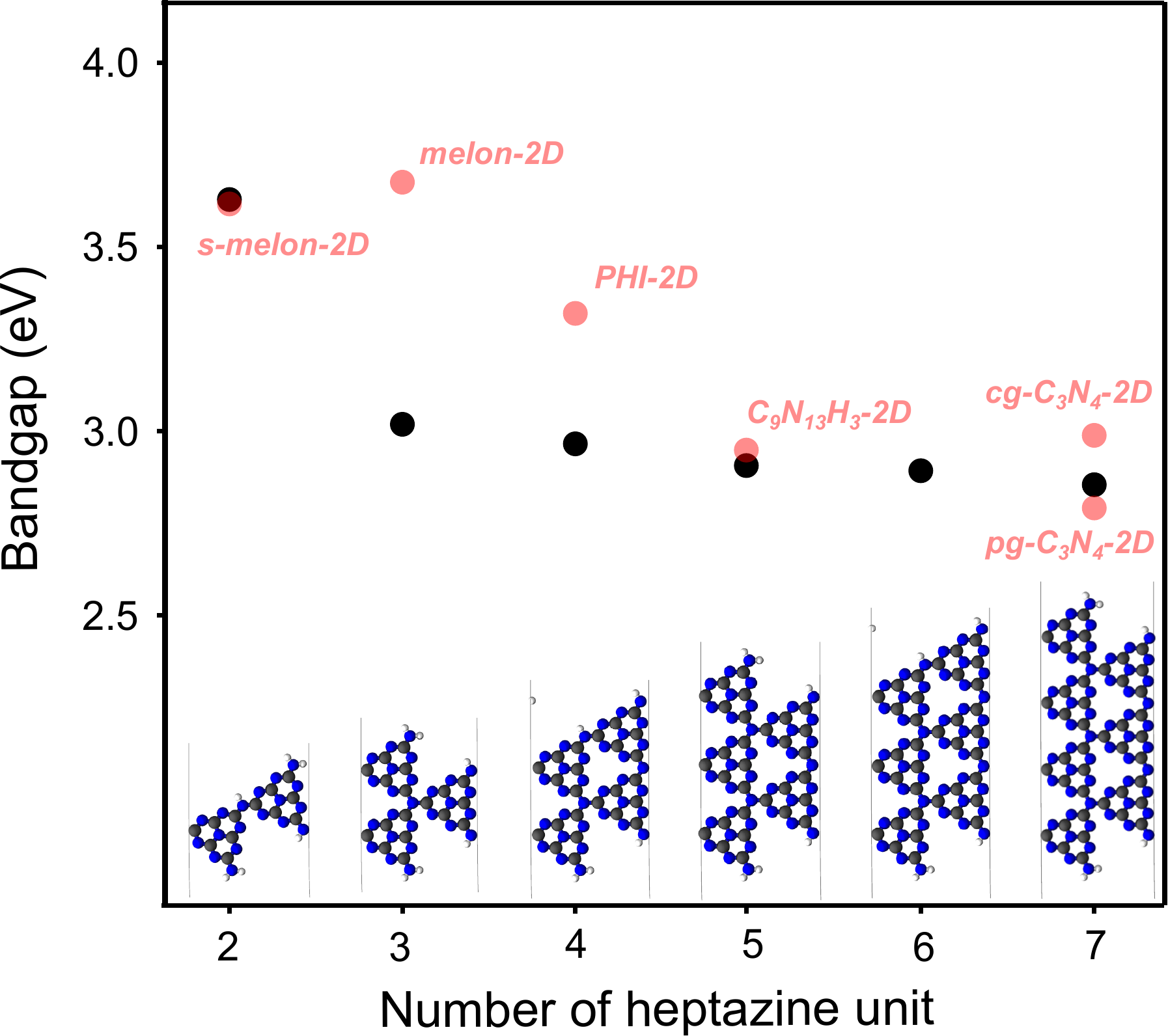}
    \caption{Calculated bandgap trends as a function of the number of heptazine monomers in hydrogen-terminated ribbon models of increasing width. Band gaps of key 2D structures are given in light red for reference (see Figure \ref{bandgap_trend}). The ribbon models neglect the influence of corrugation, see text for more details. As the number of heptazine units increases from 2 to 7, the bandgap decreases from 3.6 eV to 2.9 eV, with the most significant decrease observed already from 2 to 3 heptazine units (\textit{ca.}~0.6~eV).}
    \label{bandgap_lateral}
\end{figure}

Lastly, we studied the bandgap as function of an increasing number of heptazine units for hydrogen-terminated nanoribbons. The studied model systems, bandgap results, and a comparison to the optical properties of some key structures investigated above are given in Figure \ref{bandgap_lateral}.
Calculations show that bandgaps converge to a value close to that of the 100~\% DoC pg-\ce{C3N4}-2D model for ribbons that are 5$-$7 heptazine units wide. Already at a width of 3 heptazine units, a significant reduction of the bandgap from \textit{ca.}~3.6 to 3.0~eV is observed. This result suggests that already small areas of g-\ce{C3N4} are sufficient to produce the bandgaps observed for the fully-condensed 2D materials.

Note that corrugation is not taken into account in this test set and thus, values are closer to the flat pg-\ce{C3N4}-2D model rather than the corrugated cg-\ce{C3N4}-2D model. For larger ribbons (5$-$7 units),  it is expected that bandgap values would shift to larger values if corrugation was accounted for, analoguous to the difference between cg- and pg-\ce{C3N4}-2D.

\subsection{The Influence of Hydrogen Bonding on the melon Structure}\label{sec:The_Influence_of_Hydrogen_Bonding}

The melon-2D and melon-3D models are set apart from the other structures because their individual strings interact via hydrogen bonds exclusively rather than via any bridging covalent bonds. The investigated melon structures exhibit high thermodynamic stability both in the literature\cite{Botari2017,melissen2016dft} and our calculations, as shown in Figure \ref{phase_diagram_2D3D}. However, the formation energy trends were obtained by optimizing the atomic configurations at a temperature of 0~K; any temperature dependence of the calculated Gibbs free energy values is accounted for using the approximations described in the method section. Notably, it is known from atomistic and experimental studies on the structure of water that hydrogen bond lengths elongate quadratically as a function of increasing temperature.\cite{dougherty1998watertemperature} The calculations presented here therefore likely underestimate the separation distance between the hydrogen-bonded melon strings and overestimate their stability at ambient temperature and particularly at elevated temperature during synthesis.

\begin{figure}[htbp]
    \centering
    \includegraphics[width=\linewidth]{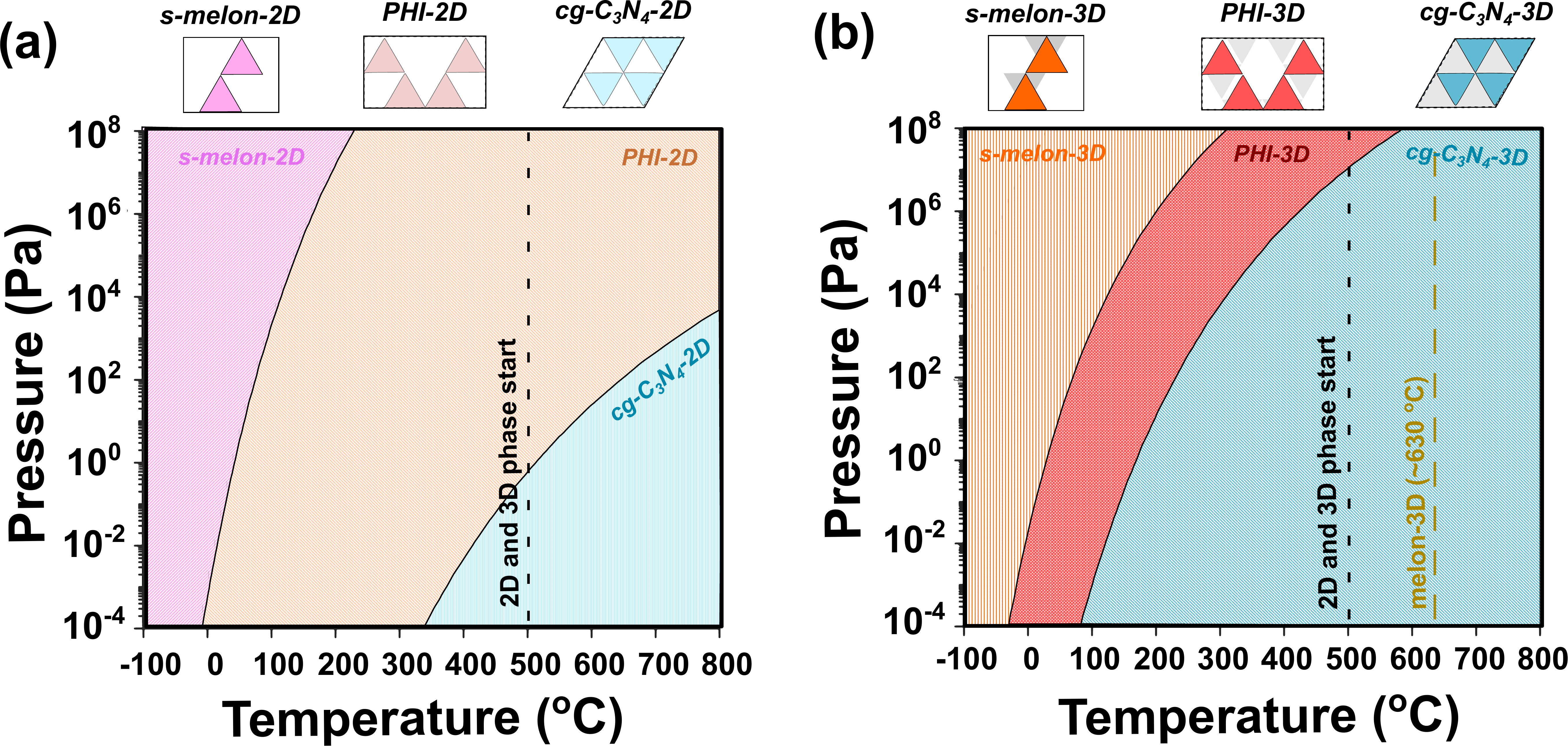}
    \caption{Formation free energy phase diagrams of 2D and 3D structures updated from Figure \ref{phase_diagram_2D3D}. The hydrogen-bonding stabilized melon-2D/-3D structure are excluded, instead the single-string s-melon-2D/-3D structures are shown. The dotted and the dashed line indicated the experimentally observed temperature where formation of 2D and 3D structures begins\cite{li2018temp,yan2009temp2,lau2022tour} and where the melon-3D structure was found to form,\cite{lotsch2007unmasking} respectively.}
    \label{phase_diagram_2D3D_renew}
\end{figure}

To account for this shortcoming, another melon-based periodic model is investigated: the single-stringed model, s-melon-2D, as presented in Figure \ref{list_of_2d3d_structures}. This structure constitutes an extreme case, \textit{i.e.} where melon strings are separated far enough that no more interaction via hydrogen bonds takes place between them. This model is therefore likely more representative for the string separation distance at high temperatures, such as during synthesis. Figure \ref{phase_diagram_2D3D_renew} compares the $T$-$p$-dependent stability of the favourable PHI and cg-\ce{C3N4} structures with the s-melon structure instead of the hydrogen-bonded melon-2D structure shown in Figure \ref{phase_diagram_2D3D}.

By comparing Figure \ref{phase_diagram_2D3D_renew} and Figure \ref{phase_diagram_2D3D}, it becomes clear that hydrogen bonding significantly contributes to the stability of the melon models. The exclusion of this contribution shifts the stability region of s-melon-2D structure in Figure \ref{phase_diagram_2D3D_renew} a) to a lower temperature and the PHI-2D structure now appears to be the most favourable configuration at reported experimental synthetic conditions as indicated by the dotted line.\cite{li2018temp,yan2009temp2,lau2022tour} In case of the 3D models, Figure \ref{phase_diagram_2D3D_renew} (b) shows that now, the cg-\ce{C3N4}-3D and PHI-3D structural motifs are in thermodynamic competition, which means that a mixture of both structure can be expected to form under typical synthetic temperature conditions (500--630 $^{\circ}$C). 

\cleardoublepage

\section{Discussion}

The purpose of this study is to elucidate the relationship between the structure of PCN materials and the corresponding optical properties from a thermodynamic point of view and as a function of material dimensionality. In the following, three key structural parameters that greatly affect the optical properties will be discussed in more detail based on the presented calculation results: (i) degree of condensation, (ii) stacking, and (iii) corrugation. 

%Higher degree of condensation and stacking increase the relative stability and reduce the optical H-L gap or bandgap, simultaneously. The corrugation also turns out to increase thermodynamic stability\cite{deifallah2008electronic,gracia2009corrugated1,melissen2015relationshipcorru3,gao2020corrugation2}, whereas enlarges the optical H-L gap or bandgap than that of the planar structure. From results of the relative thermodynamical stability and the trend of the optical gap, we not only provide insight into the structure-optical relationship but also  propose the reaction mechanism under the realistic synthetic conditions.

\subsection{Correlation between Degree of Condensation and Synthetic Conditions} \label{interplay} % need to give more weight to "DoC"
\begin{figure}[ht]
    \centering
    \includegraphics[width=\linewidth]{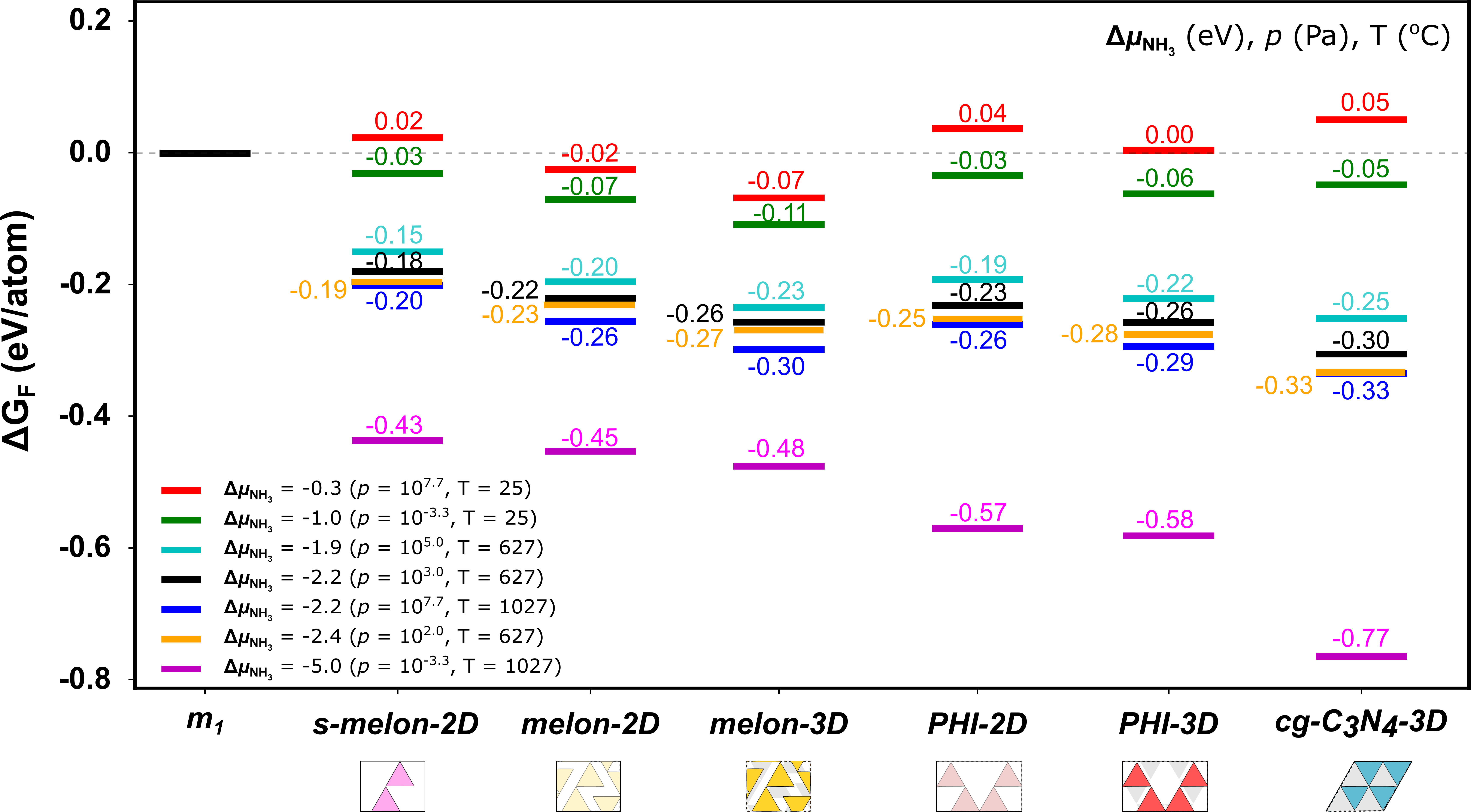}
    \caption{Free energy of formation for key 2D and 3D structures as a function of a selected set of $\mu_\text{\ce{NH3}}$ conditions. The free energy of formation is normalized to eV/atom to be able to compare structures with different total number of atoms.}
    \label{formation_free}
\end{figure}

This section discusses in detail how different synthetic conditions can facilitate different degrees of condensation (DoC) of the synthesized material, according to the calculated formation energy trends. Figure \ref{formation_free} summarizes the formation free energy trends for the most stable structures presented in Figures \ref{phase_diagram_molecules}, \ref{phase_diagram_2D3D}, and \ref{phase_diagram_2D3D_renew}. An unfiltered overview containing all structures is shown in Figure \ref{formation_free_overall_si}. In this depiction, emphasis is placed on highlighting the change of trends of the thermodynamically preferred structure as a function of $\Delta \mu_\text{\ce{NH3}}$, \textit{i.e.} at various $T$ and $p$ conditions.

The formation free energy, $\Delta G_\text{F}$, is calculated for each structure from \ce{NH3} and melamine as the educts. The middle-ranged cyan, black, yellow, blue, and orange lines represent the $T$-$p$ parameter space that can be achieved under typical laboratory conditions. 

From the formation energy trends, it is evident that the DoC is proportional to $\Delta \mu_\text{\ce{NH3}}$; structures with a higher DoC become more favourable as $\Delta \mu_\text{\ce{NH3}}$ becomes more negative. In the aforementioned middle range (blue to cyan lines), all structure show similar thermodynamic stability. We can therefore conclude that from a thermodynamic perspective, the experimentally synthesized material most probably is a mixture of these different structural motifs with no strong preference for any one particular motif.

In addition, the results suggest that certain structural motifs may become dominant when specific synthetic conditions are enforced. On one extreme end of the $T$-$p$ spectrum, we predict that the less-condensed melon string structures can be enforced by performing the synthesis at low temperature conditions and high $p_\text{\ce{NH3}}$, for example in a closed reaction vessel (red and green lines). On the other extreme end, if the synthesis is performed at high temperature and low $p_{\ce{NH3}}$, for example in a reaction vessel with an applied vacuum (magenta line), the more condensed cg-\ce{C3N4} and PHI structures become favourable. While these extreme cases are not representative of conditions that can be readily achieved in the laboratory, they serve to illustrate the influence of the parameters of temperature and pressure on the expected composition of the PCN material. 

Therefore, in order to obtain a PCN with a high DoC and a optical gap of 3~eV or less, these computational trends suggest that high temperature and low $p_{\ce{NH3}}$ conditions should be maintained during the synthesis. 
These calculated trends are corroborated by experiments by Inoki \textit{et al.}\cite{inoki2020lowpressure}, who reported successful synthesis of a g-\ce{C3N4} structure under controlled low $p_{\ce{NH3}}$ (\textit{ca.} $9.0~\times~10^{3}$ Pa) at 550 $^\circ$C. Their synthesized material shows a bandgap of 2.03 eV with an average crystal size of less than 3~nm. On the other hand, Lotsch \textit{et al.} showed that formation of the melon structural motif can be promoted by conducting synthesis in a closed ampule at 630 $^\circ$C.\cite{lotsch2007unmasking} Further condensation of the melon strings is likely hindered by the relatively high \ce{NH3} $p_{\ce{NH3}}$ within the ampule. Both of these experimental observations are in good agreement with the predicted trends presented in this work (Figure \ref{phase_diagram_2D3D}(d)).

Finally, the presented results may help to reconcile conflicting observations from theory and experiments. While published theoretical results predict that high DoC is a requirement to achieve the desired bandgap, experiments measure bandgaps of 2.6--2.7~eV for materials where elemental analysis shows high H content that would indicate a low DoC.\cite{lotsch2006triazines,liu2016graphitic,lau2022tour} According to the presented trends, the synthesized material likely contains a mixture of different more or less condensed structural motifs, giving rise to desirable optical properties in g-\ce{C3N4}-like domains embedded in a matrix of less-condensed PCN  such as the melon string motif. Consequently, this conclusion raises a follow-up question: how large do these highly-condensed domains need to be in order to exhibit the desired properties? This question will be addressed in section \ref{stacking}.

\subsection{Interconversion} \label{interconversion}

It is clear from the presented computational trends that there is thermodynamic competition between the melon-3D structure and the idealized, fully-condensed cg-\ce{C3N4}-3D structure for the end product. Consequently, the question arises if and how interconversion can take place between these structure. More specifically, in this section we discuss whether a thermodynamic pathway exists to convert these phases into each other, for example by annealing the synthesis product under specific conditions.

\begin{figure}[htbp]
    \centering
    \includegraphics[width=\linewidth]{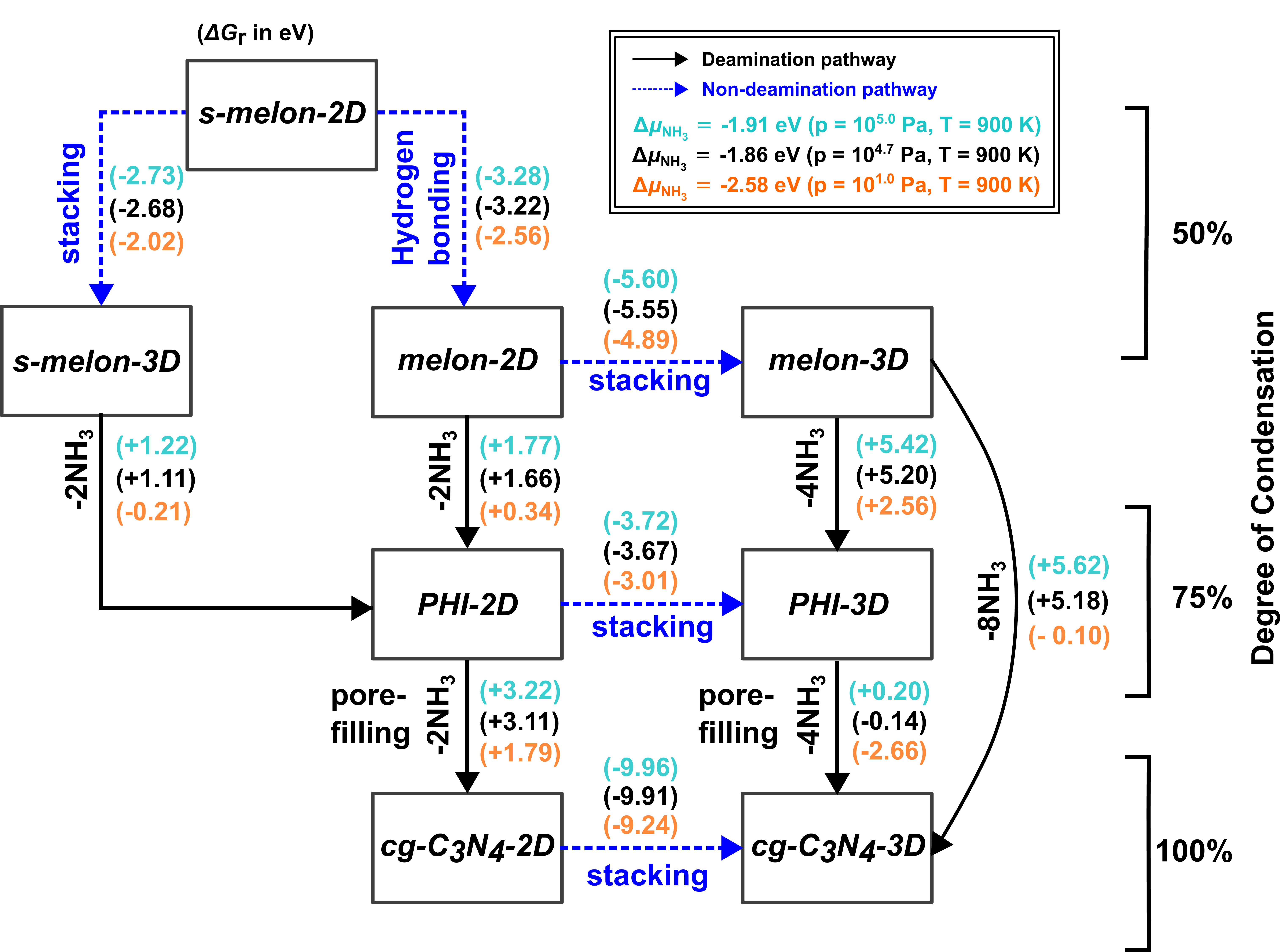}
    \caption{Reaction flow chart connecting structures via the calculated reaction free energy, $\Delta G_\text{r}$, which depends on $\Delta\mu_\text{\ce{NH3}}$ (-2.6 eV $<$ $\Delta\mu_\text{\ce{NH3}}$ $<$ -1.9 eV). Temperature is fixed to 628 $^{\circ}$C, which constitutes the maximum temperature before carbonization of the material sets in, so that the reaction free energy trends depend solely on $p_{\ce{NH3}}$. Two reaction mechanisms are suggested: i) The non-deamination pathway involving stacking and hydrogen bonding and ii) the deamination pathway involving release of \ce{NH3}. The structures are classified by their degree of condensation (DoC). The "pore-filling" route is suggested to connect PHI structure to cg-\ce{C3N4} structure; see text for more details.}
    \label{reaction_free}
\end{figure}

From a structural point of view, direct conversion from melon to the fully-condensed structure is not straightforward (DoC = 50\% for melon, DoC = 100\% for cg-\ce{C3N4}). It is therefore reasonable to assume that the PHI structure (DoC = 75\%) may be an intermediate structure connecting them. In fact, the PHI structure has already been successfully synthesized, indicating that this structure is at least metastable, if not thermodynamically favourable under specific conditions.\cite{doblinger2009structure,dontsova2015triazoles,savateev2017potassium,chen2017easier,lin2018crystalline,schlomberg2019PHI} We therefore hypothesize that the large pores in the PHI structure may be filled with heptazine monomers under certain conditions to give the cg-\ce{C3N4} structure. Selected hypothetical pathways and their associated reaction free energy trends at different $T$-$p$ conditions are summarized in Figure \ref{reaction_free}.
More detailed $p$-$T$ phase diagrams from which these trends were summarized are shown in Figure \ref{reaction_free_si}.

Here, the reaction free energy $\Delta G_\text{r}$ between any product and an educt state is calculated as
\begin{eqnarray}
    \Delta G_\text{r}(T,p) = n\ G^{\, \text{product}}(T, p) - m\ G^{\, \text{educt}}(T, p)\ - o\ G^\text{\ce{NH3}}, %\label{}, \eqref{}
\end{eqnarray}
where $G^{\, \text{product}}(T, p)$, $G^{\, \text{educt}}(T, p)$, and $G^{\text{\ce{NH3}}}(T, p)$ constitute the Gibbs free energy of the product and educt configurations as well as the \ce{NH3} molecule, respectively, and $n$, $m$, and $o$ are stoichiometric factors.

The trends shown in Figure \ref{reaction_free} were calculated at 628 $^{\circ}$C, which is close to the decompostion temperature of \textit{ca.}~630 $^{\circ}$C.\cite{yan2009temp2} The reaction profile can then be controlled by $p_{\ce{NH3}}$. The proposed reaction profile conveys two important thermodynamic factors that determine the end structure at the respective conditions:
\begin{enumerate}
    \item The non-deamination pathway is thermodynamically more favourable than the deamination pathway because stabilization occurs via hydrogen bonds and/or stacking. However, this process does not increase the DoC.
    \item $p_{\ce{NH3}}$ plays a pivotal role in controlling the reaction free energy of the deamination pathways, which increases the DoC.
\end{enumerate}
From these trends, it is clear that the melon structure represents a thermodynamic sink. The conversion from melon-3D to the idealized cg-\ce{C3N4}-3D structure only becomes exothermic at drastic conditions of $p_{\ce{NH3}} = 10$~Pa. The same is true for the deamination pathway leading from s-melon-3D to PHI-2D. However, if the PHI-2D stage can be reached, which will likely be in equilibrium with the melon-2D structure due to similar formation free energy trends from s-melon-2D, the subsequent stacking to PHI-3D and pore filling reaction to cg-\ce{C3N4}-3D, which already becomes spontaneous close to atmospheric pressure, are significantly downhill.

To develop a conversion strategy, it is crucial to determine from which point in this flowchart one starts. For example, if synthesis of a PHI material has already succeeded, for example with the help of templating effects,\cite{chen2017easier,savateev2017potassium} further condensation due to pore filling is possible at comparatively mild conditions. However, converting a melon-3D material into cg-\ce{C3N4}-3D may require drastic conditions. When synthesis starts from the educts, a mixture of melon-3D and more condensed material, as signified by the idealized cg-\ce{C3N4}-3D system, can be expected due to thermodynamic competition of the s-melon-2D $\rightarrow$ s-melon-3D and the s-melon-2D $\rightarrow$ melon-2D pathway.

\subsection{Influence of Stacking \textit{vs.} Lateral Condensation} \label{stacking}
%Stacked 2D carbon nitride is stabilized further by in-plane shifting of the layers in the crystal, thereby reducing the repulsive forces resulting from the $\pi-\pi$ interlayer interactions.\cite{fina2015structural}. Indeed, it can be confirmed by the fact that the peak mainly found in XRD corresponds to the interlayer spacing at 27$\circ$ in general synthesis condition.\cite{thomas2008graphitic,tyborski2013crystal,fina2015structural}
The presented results show that stacking, \textit{i.e.} the 2D $\rightarrow$ 3D transition, is thermodynamically more favorable than condensation or pore filling reactions. Furthermore, a central conclusion of sections \ref{interplay} and \ref{interconversion} is that the synthesized material is most likely a mixture of structures, particularly of the melon-3D structure and partially pore-filled PHI-3D. As mentioned at the end of section \ref{interplay}, we now face an important question: how big do these highly-condensed domains, which are likely interspersed throughout a host matrix of less strongly condensed melon- or PHI-type structures, need to be in order for the material to exhibit the bandgaps commonly measured in experiments?

In order to investigate this question, we performed a convergence study for strongly condensed carbon nitride nanoribbons of increasing width, see Figure \ref{bandgap_lateral}. Results show that the optical gap of a ribbon structure composed of 7 or more heptazine units ($E_\text{gap} = 2.85$~eV) has reached that of the fully-condensed structures (cg-\ce{C3N4}-2D: 2.99~eV, pg-\ce{C3N4}-2D: 2.79~eV). This result underlines that small domains of strongly-condensed PCN of \textit{ca.}~3--4~nm$^2$ can already exhibit the desired optical properties. Even though our computational predictions show that the material prefers to grow via stacking rather than laterally, condensation of only 7 heptazine units should be readily possible. This extrapolated minimum crystalline size is in good agreement with experiments by Inoki \textit{et al.}, who calculated a bandgap of 2.06 eV for PCN crystals of \textit{ca.}~3~nm size.\cite{inoki2020lowpressure}

Furthermore, the average HOMO-LUMO gap of all studied tetramers was computed to be 3.61 eV, which is comparable to that of the melon 2D structures (3.62 eV and 3.68 for s-melon-2D and melon-2D, respectively; see Table \ref{table:table_optical_gap}). While the absolute values of molecular HOMO-LUMO gaps and bandgaps from periodic calculations are not directly comparable, this trend is still noteworthy. It is also corroborated by a combined experimental$-$theoretical study by Li \textit{et al.}, who report that the optical gap of heptazine oligomers rapidly converges with size.\cite{li2018temp} Therefore, this result not only emphasizes again that a high DoC is a key driving force for red shifting of the optical gap, but is also in a good agreement with the ribbon model convergence study which shows that only small domains of high condensation PCNs are required to produce a desirable bandgap.

\subsection{Influence of Corrugation}

Corrugation of carbon nitride materials is a significant factor in terms of thermodynamic stability and optical properties. Using GGA-DFT and MP2 methods, Gracia and Kroll found that corrugation enhances the thermodynamic stability of condensed carbon nitride structures by relieving the repulsive interaction between adjacent lone-pair electrons of the nitrogen atoms.\cite{gracia2009corrugated1} This conclusion is in agreement with many other published studies\cite{deifallah2008electronic,gracia2009corrugated1,melissen2015relationshipcorru3,gao2020corrugation2} as well as the simulation results presented here. For example, our calculations show that the corrugated fully-condensed cg-\ce{C3N4}-2D model is more stable than its fully flat counterpart pg-\ce{C3N4}-2D ($\Delta E = 0.7$~eV/heptazine unit). In terms of optical properties, corrugation increases the bandgap, see Table \ref{table:table_optical_gap}. For example, the flattened pg-\ce{C3N4}-3D structure has a 0.6~eV narrower bandgap compared to the corrugated --- and thermodynamically more stable --- cg-\ce{C3N4}-3D structure.

Out-of-plane corrugation of the material occurs in order to minimize repulsive interactions between adjacent N atoms.\cite{gracia2009corrugated1} Here, structural factors contributing to corrugation are examined in more detail, namely the DoC, lateral crystallite size, and stacking. To this end, we compare the calculated root-mean-square deviations (RMSD) of the out-of-plane shift of atomic coordinates for different 1D, 2D, and 3D structures, see Table \ref{table:table_corrugation}. 

First, lateral crystallite size is examined. The RMSD shift increases from 0.01 \AA\ for the heptazine monomer (h$_1$) up to 0.25 \AA\ for the pentamer (h$_{51}$). This result suggests that larger structures corrugate more easily. Section \ref{interplay} already established that the experimental material is most likely a mixture of different structural motifs and section \ref{stacking} established that already small areas of g-\ce{C3N4} can exhibit bandgaps of 3~eV and less. In light of these trends, the present results suggest that in order to achieve a desirable bandgap, a material with only small connected patches of the same structural motif, thereby avoiding corrugation, is desirable. A recent study by Mitoraj \textit{et al.} has used this principle to good effect.\cite{mitoraj2021study} The group employed functionalization with a small fraction of azo-linkers in the inner domain to trigger local flattening of the material, thereby redshifting the absorption spectrum.

The DoC can have an even more profound impact on corrugation compared to crystallite size. For example, comparing RMSD shifts for the melon-2D (0.00 \AA), PHI-2D (0.57 \AA), and cg-\ce{C3N4}-2D (0.80 \AA) structures in Table \ref{table:table_corrugation}, it becomes apparent that corrugation increases as the DoC increases, which counteracts the trend of narrowing bandgaps with increasing DoC (see Table \ref{table:table_optical_gap}). This result once again emphasizes that if the goal is to obtain a PCN material with low bandgap, connected areas of the same structural motif should be kept small since corrugation becomes more pronounced as these areas increase in size.

Finally, stacking appears to inconsistently affect the RMSD shift of the atomic coordinates. For example, the out-of-plane RMSD shifts of atoms in the 2D and 3D cg-\ce{C3N4} models are close to identical (see Table \ref{table:table_corrugation}). The bandgaps of these models are close to identical, see Table \ref{table:table_optical_gap}. On the other hand, the bandgap of the fully-flattened pg-\ce{C3N4} model is reduced by 0.44~eV due to stacking. Going from PHI-2D to PHI-3D, the extent of corrugation decreases, thereby slightly reducing the bandgap. This result indicates that at least based on the present data set, there is no obvious systematic trend connecting stacking and corrugation. Therefore, any new model system requires detailed investigation in this regard. 

To summarize, corrugation has a profound impact on the stability and electronic structure of carbon nitride materials. In calculations, corrugation is found to stabilize structures thermodynamically and to increase the bandgap. The latter point is two-pronged: corrugation weakens the $\pi$-$\pi$ interactions in the lateral direction but also weakens interlayer interactions perpendicular to the layers. From these results, we hypothesize that if the goal is to achieve a material with a small bandgap, synthetic conditions should be optimized to yield a PCN consisting of small domains of the same structural motif since corrugation becomes more pronounced with increasing domain size. As discussed in section \ref{stacking}, small domains of g-\ce{C3N4} material interspersed in a host matrix of less-condensed material are sufficient to obtain a desirable bandgap.

\subsection{Postulation of a More Complete Computational Carbon Nitride Model} \label{postulated}

In this final section, we aim to combine the insights gained throughout this study to postulate a more complete model for the computational investigation of PCNs. While the experimental material is arguably too complex to be treated with any one separate model, we are convinced that there is nevertheless great value in taking steps towards representing as many of the intricacies of this fascinating material class as possible in simulations.

\begin{figure}[htbp]
    \centering
    \includegraphics[width=\linewidth]{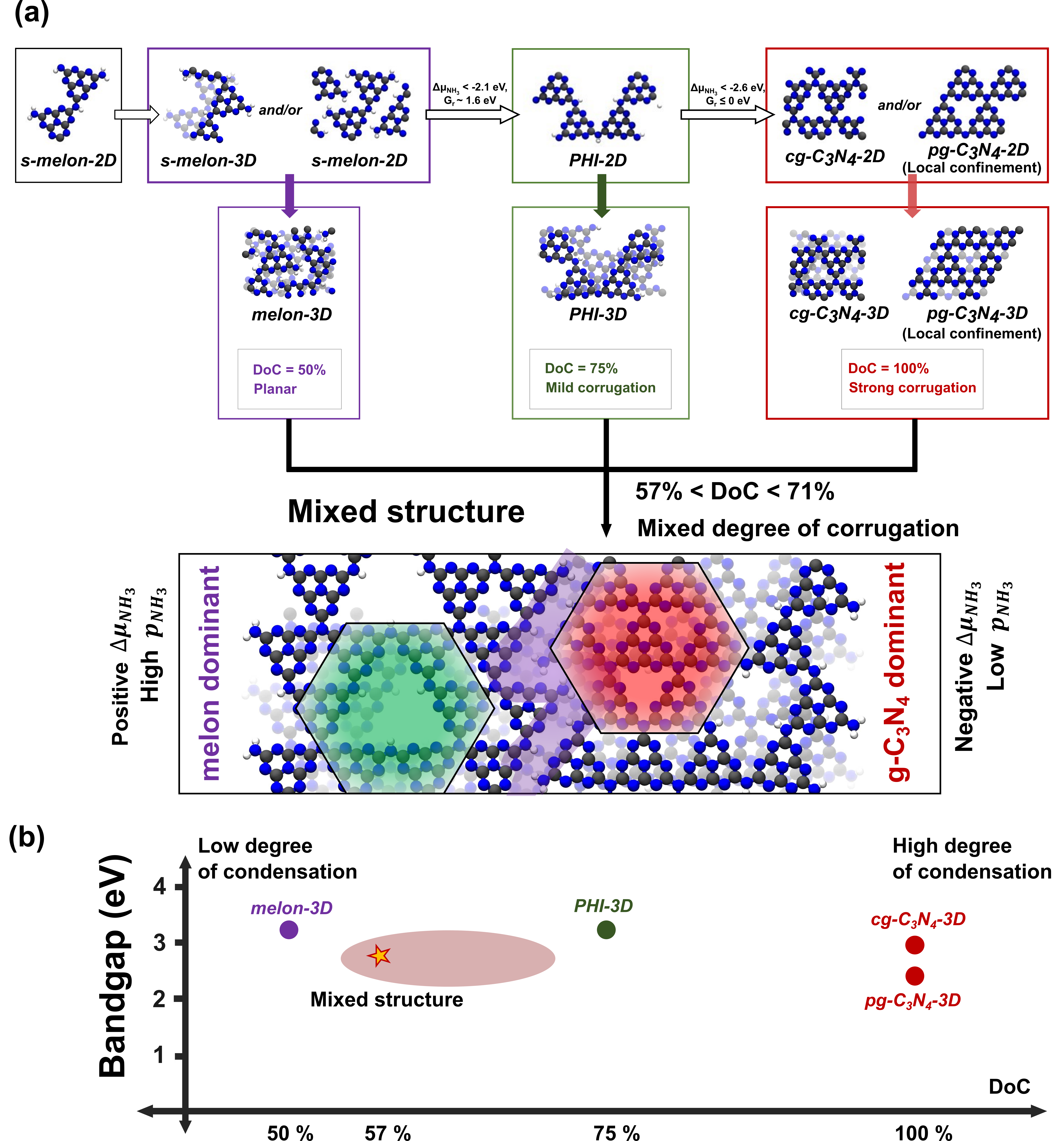}
    \caption{(a) Illustration of the postulated "mixed" PCN model as well as the parent structures it is composed of. (b) Calculated bandgaps for the mixed structural motif as well as for the parent structures as a function of the DoC. The oval shape indicates the range of DoCs and bandgaps obtained for the four tested mixed models; the yellow star indicates the most promising model that is also depicted in (a).}
    \label{proposed_structure}
\end{figure}

The new model, herein referred to as the "mixed" model, should satisfy the following conditions:
\begin{enumerate}
    \item The melon structure (50\% DoC) should be represented in the model as it constitutes one of the thermodynamically most stable structures found in the present study.
    \item The PHI model (75\% DoC) is another thermodynamically stable candidate structure that warrants representation, also because successful synthesis of this motif has been reported using a templating method with salt melts.\cite{bojdys2008ionothermal,savateev2017towards,schlomberg2019PHI}
    \item From bandgap results in this study, only the fully condensed cg- and pg-\ce{C3N4} structures can sufficiently explain the experimental bandgap. The planarized pg-\ce{C3N4}-3D structure gives a bandgap as low as 2.35 eV. Calculated free energy trends suggest that these structures form under low $p_\text{NH3}$ conditions. This 100\% DoC structure should therefore be represented as well.
    \item Results on the influence of corrugation as well as the presented lateral size \textit{vs.} bandgap data show that small domains of g-\ce{C3N4} with low corrugation are sufficient to achieve a narrower bandgap. The model size, therefore, can remain overall small, with rapid alternations between the above structural motifs.
\end{enumerate}
The resulting model as well as a comparison between its DoC and optical properties with those of its three parent structures is illustrated in Figure \ref{proposed_structure}.
Shown here is only the most promising mixed model with 57\% DoC. Three other structures with DoC up to 71~\% were investigated as well and are shown in the Supporting Information, see Figure \ref{list_of_mixed_structures} for illustrations of their structures and Table \ref{table:table_optical_gap} for the optical properties.
%add limitation of RMSD since the corrugation can generate a lot of local minima state
%The lower DoC of the mixted models than that of PHI models 
 
%In the $p$-$T$ space, the proposed mixed structure shows a similar thermodynamic stability region as the PHI-3D and cg-\ce{C3N4}-3D structures. 
Because the fully-condensed structural motif (red hexagon) is confined in a small domain, only mild corrugation occurs (0.573 \AA\ \textit{vs.} 0.804 \AA\ for cg-\ce{C3N4}) and the resulting optical bandgap is \textit{ca.}~2.9~eV, see Table \ref{table:table_optical_gap}. It is noteworthy that the optical bandgap of the mixed structures is lower than that of cg-\ce{C3N4}-3D, likely due to the aforementioned encapsulation and local flattening. This result confirms the initial hypothesis that extended condensed structures are not necessary to achieve a desirable bandgap, and highlights that mixed structures can easily give rise to low bandgaps.

To the best of our knowledge, this is the first time that a model system capturing this many important facets of PCNs at once is presented in literature. This study, however, is limited to the structure of the pure PCN materials and does not take into account functional end groups such as --\ce{N}--\ce{H} and --\ce{N}--\ce{CN} which may be able to further adjust the local electronic structure.\cite{schwarzer2013endgroup} The influence of functional groups on electronic properties and reactivity will be explored in the future using this model system as basis.

\section{Conclusion}
The thermodynamic stability and the optical properties of heptazine-based carbon nitride materials were investigated using accurate hybrid density functional theory calculations. Three structural parameters were identified that strongly affect the optical properties: 
\begin{enumerate}
    \item degree of condensation (DoC; decreases bandgap and decreases stability),
    \item stacking (decreases bandgap and increases stability), 
    \item and corrugation (increases bandgap and increases stability).
\end{enumerate} 
Materials with a high DoC also show pronounced corrugation, which presents a dilemma if material optimization towards narrow bandgaps is the goal. Materials with high DoC and pronounced corrugation form at conditions where the chemical potential of \ce{NH3}, $\mu_{\ce{NH3}}$, is strongly negative, \textit{i.e.} at high temperature and low $p_{\ce{NH3}}$. The melon string structure is preferred at less negative $\mu_{\ce{NH3}}$ but exhibits a large bandgap. We identified interconversion pathways starting from the melon string structure towards more condensed structures. However, from the relatively small differences in thermochemical stability between these structures, it is expected that materials synthesized under typical experimental conditions consist of a mixture of more and less condensed structural motifs. The thermodynamic preference for stacking over lateral condensation likely further discourages formation of extended areas of the same motif.

From the calculated trends, we postulate a complex computational model that combines features of the melon string, poly(heptazine imide) (PHI), and g-\ce{C3N4} model systems. This mixed structure shows a similar thermodynamic stability region as its parent molecules and exhibits, despite its lower DoC of 57\%, a narrower bandgap of 2.9 eV compared to the fully condensed but strongly corrugated cg-\ce{C3N4}-3D model. The bandgap is likely a result of local flattening of the small, encapsulated, strongly-condensed motif featured in the model.

% the sentence below is just my trial to change it in a different way.. as Radim commented on it. Technically nothing has been changed lol. cheers!
% The present study, therefore, illustrates that comparatively small g-\ce{C3N4} domains embedded within a less condensed motifs like the melon string or PHI structures not only are thermodynamically well accessible but also can provide optical properties that are typically found experimentally.
The present study therefore illustrates that the experimentally well established favorable optical properties of PCNs can emerge from comparatively small g-\ce{C3N4} domains embedded in a framework of thermodynamically more stable and less condensed motifs like the melon string or PHI structures. In future studies, the presented computational model will be used to explore the effects of functionalization and templating ions on the stability, catalytic activity, and optical properties of PCNs.

\begin{acknowledgement}
This work was funded by the Deutsche Forschungsgemeinschaft (DFG — German Research Foundation) through TRR 234 CataLight (project no. 364549901) as well as JA 1972/27-1 (project no. 428764269). Computational resources were provided by the state of Baden-Württemberg through bwHPC and the German Research Foundation (DFG) under Grant No. INST 40/467-1 FUGG. C.I. acknowledges the German Academic Exchange Service (DAAD, Ref. No. 91676720). I.K. acknowledges the support of the Alexander von Humboldt Foundation through the Humboldt Research Fellowship.

\end{acknowledgement}


\cleardoublepage
\begin{suppinfo}
\section{Supporting information}
\setcounter{figure}{0}
\setcounter{table}{0}
\renewcommand{\thefigure}{S\arabic{figure}}
\renewcommand{\thetable}{S\arabic{table}}

\begin{figure}[]
    \centering
    \includegraphics[width=\linewidth*2/3]{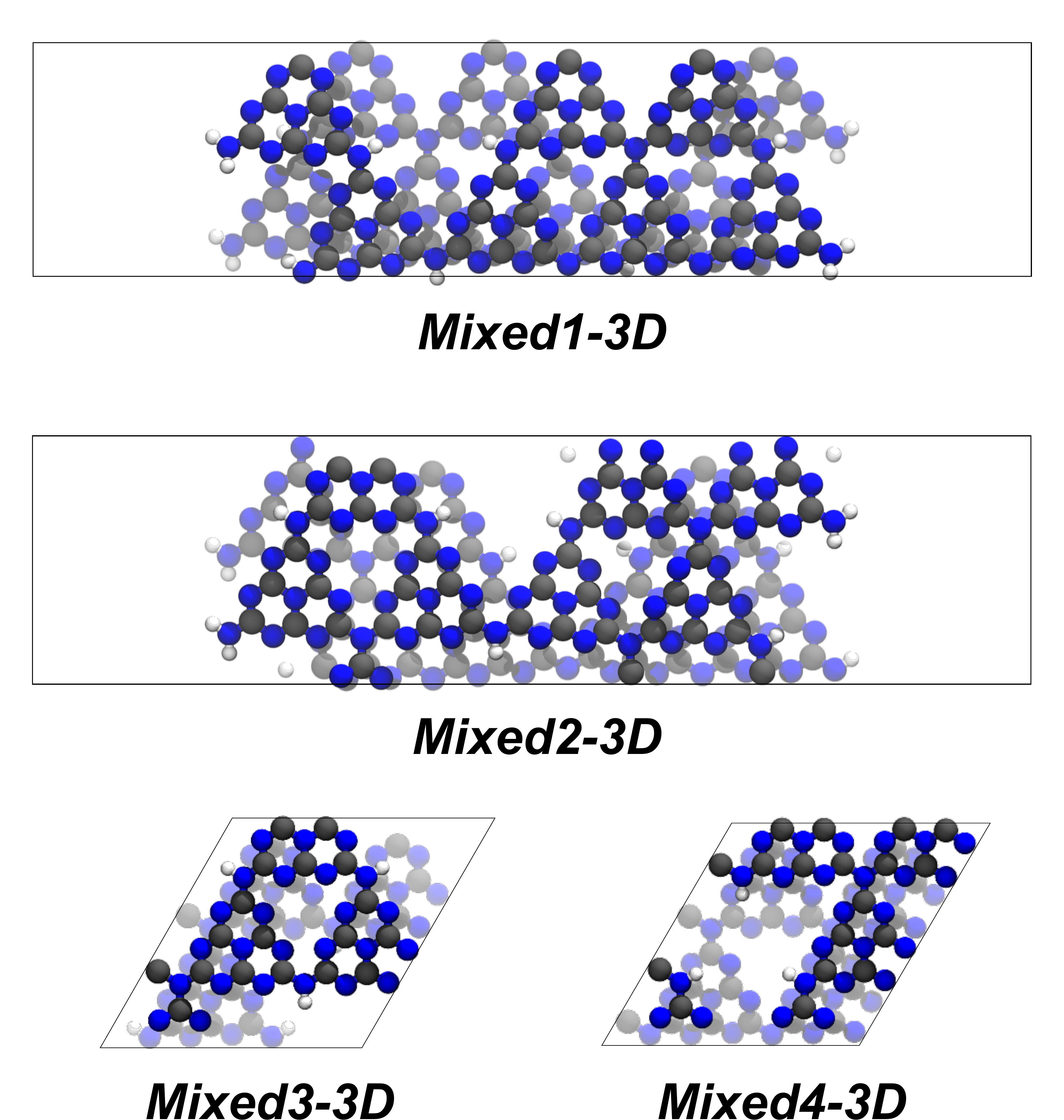}
    \caption{Proposed structures of the mixed motifs used in the study. The structures are composed of the PHI and cg-\ce{C3N4} structure because of their high DoC (79\%--94\%). The structures of Mixed1-3D and Mixed2-3D have the edge boundaries which reduce the corrugation. The both Mixed1-3D and Mixed2-3D are distinct from the sequence of the structural motifs. The Mixed3-3D is made of inclusion of PHI structure within cg-\ce{C3N4} domain to both layers and the Mixed4-3D is constructed of cg-\ce{C3N4} layer (upper) and PHI layer (bottom), respectively.} 
    \label{list_of_mixed_structures}
\end{figure}

\begin{figure}[]
    \centering
    \includegraphics[width=\linewidth*2/3]{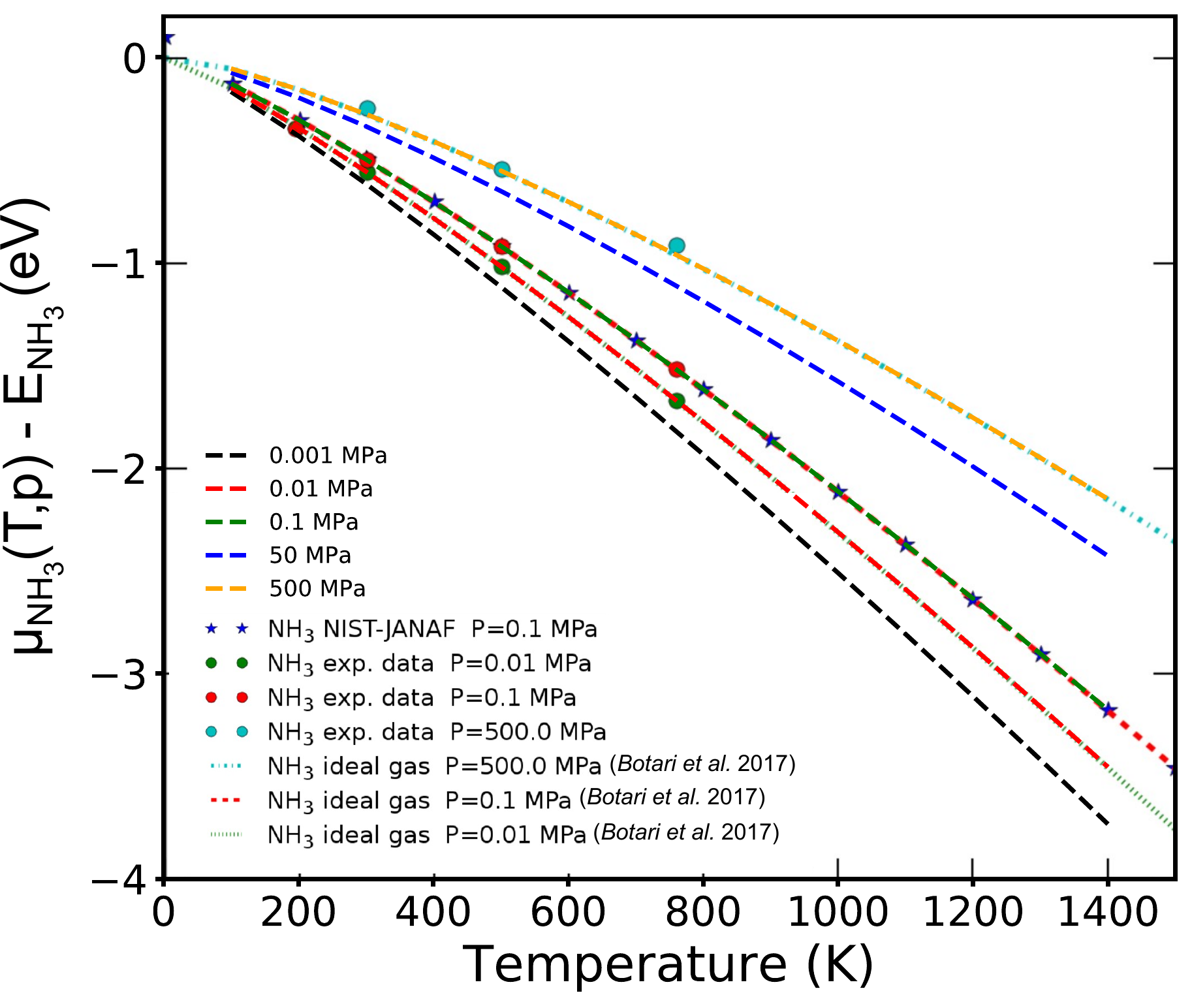}
    \caption{The evaluation of the calculated chemical potential of ammonia used in this study. The reference plots are taken from the NIST-JANAF\cite{chase1998nist,NISTJANAF} and the literature.\cite{haar1978thermodynamic,Botari2017} The dotted points ($\ce{NH3}$ NIST-JANAF and $\ce{NH3}$ exp. data) and small dotted lines ($\ce{NH3}$ ideal gas) are taken from the experimental and the literature, respectively. The bold dashed line illustrates the calculated $\Delta \mu_\text{\ce{NH3}}(\text{T},p)$ in this study. Then we added two $p_{\ce{NH3}}$ conditions of 0.001 MP and 50 MP to further trace the variations.}
    \label{chempo_dNH3}
\end{figure}

\begin{figure}[]
    \centering
    \includegraphics[width=\linewidth*2/3]{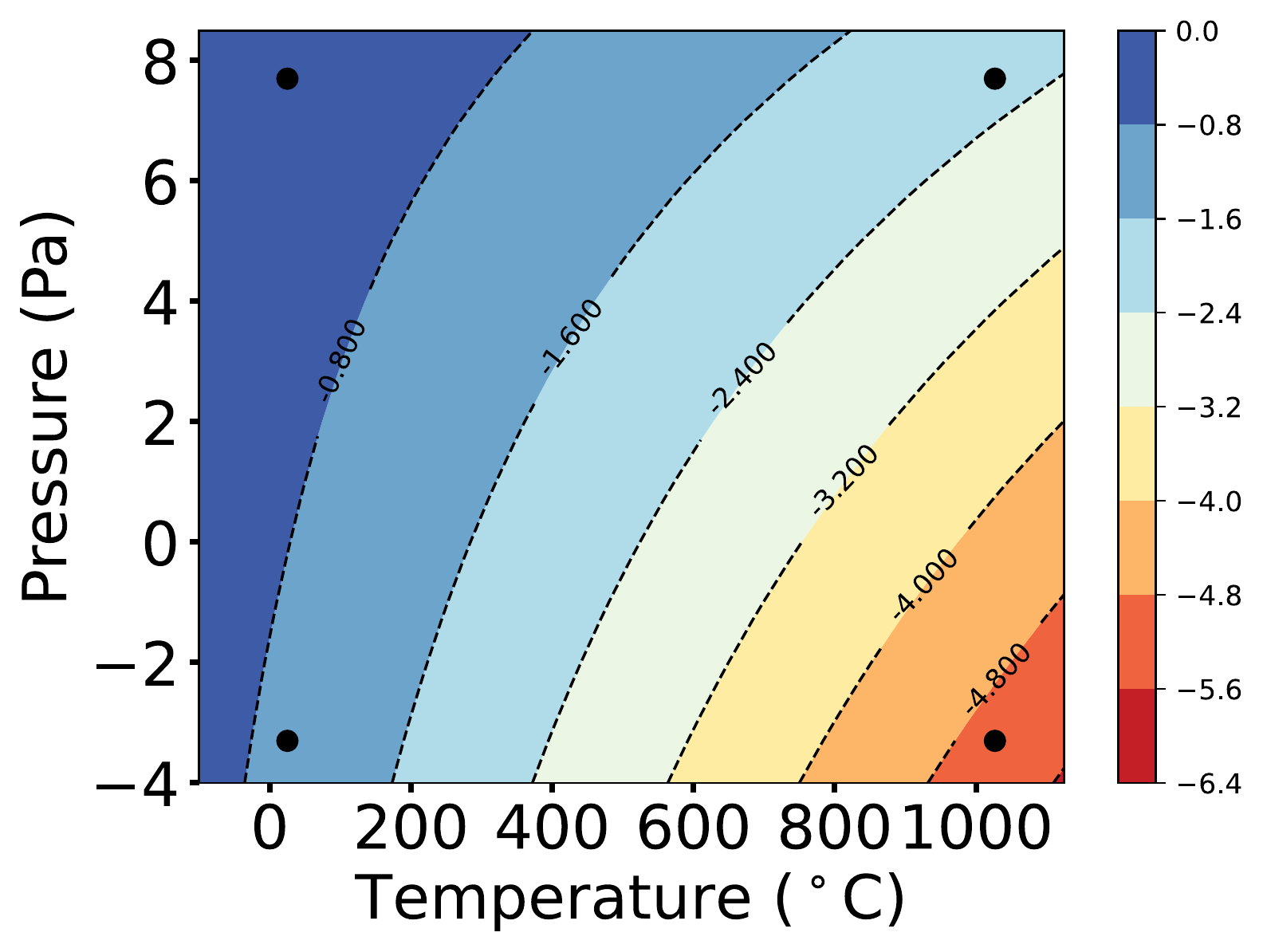}
    \caption{The contour plot of $\Delta \mu_\text{\ce{\ce{NH3}}}$. The plot is derived from the equation (\ref{equchempoNH3}). The black points is sampled the respective $\Delta \mu_\text{\ce{\ce{NH3}}}$. (See Figure \ref{formation_free} of red, green, blue, and purple.}
    \label{chempomap_dNH3}
\end{figure}

\begin{figure}[]
    \centering
    \includegraphics[width=\linewidth*2/3]{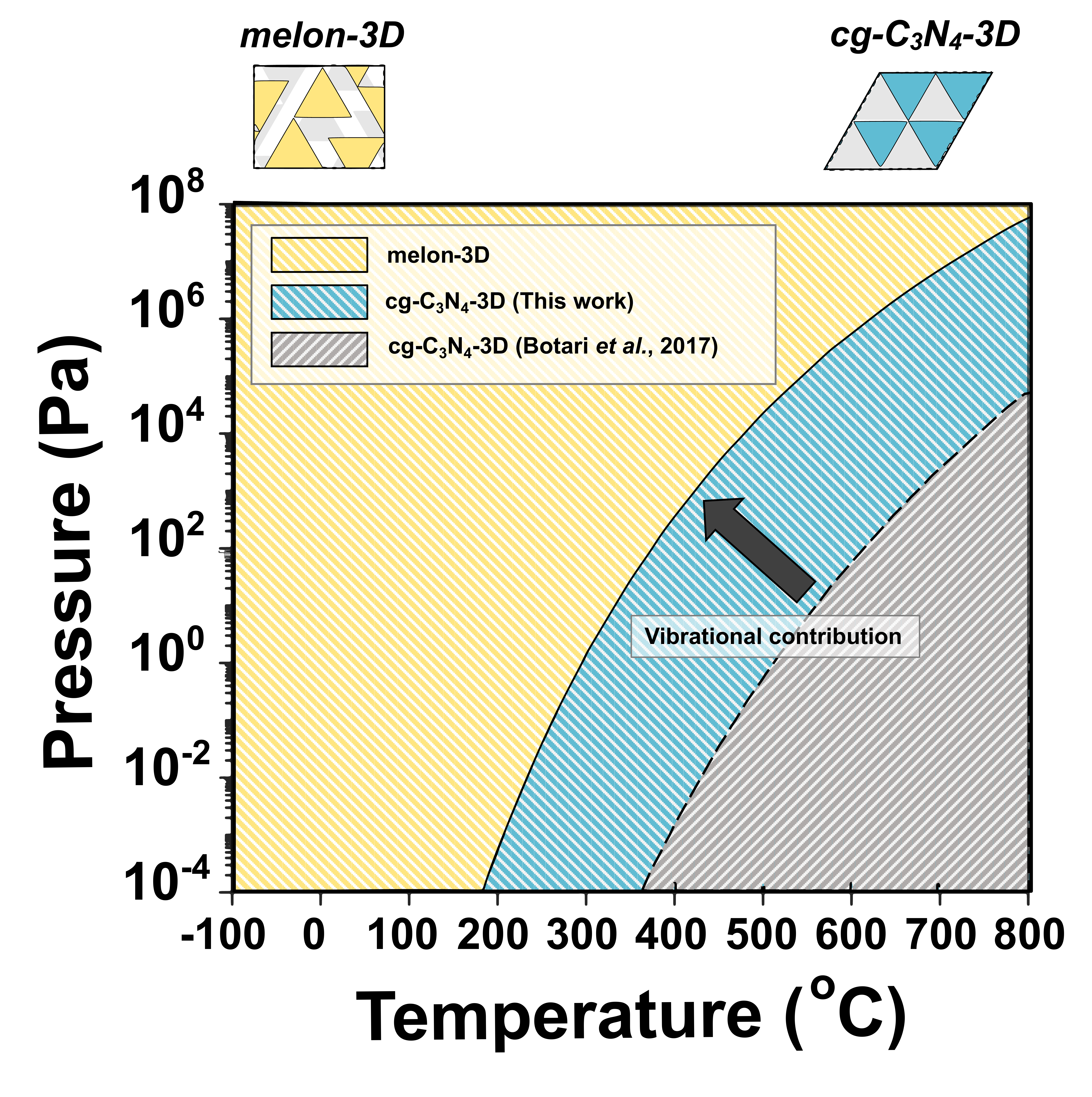}
    \caption{The calculated phase diagram of the normalized free energy of formation between melon-3D and cg-\ce{C3N4}-3D. The inclusion of the vibrational contributions increases the relative stability of of cg-\ce{C3N4}-3D (blue area) to $\Delta \mu_\text{\ce{\ce{NH3}}}$ direction. The phase diagram of melon structure from the literature is shown in gray area.\cite{Botari2017}}
    \label{example_phase_update}
\end{figure}

%${m}_{1}$	& 6.036 \\
%${m}_{2}$	& 5.43 \\
%${m}_{31}$	& 5.058 \\
%${m}_{32}$	& 4.259 \\
%${m}_{33}$	& 5.26 \\
%${m}_{34}$	& 5.193 \\
%${m}_{41}$	& 4.661 \\
%${m}_{42}$	& 5.121 \\
%${m}_{43}$	& 4.752 \\
%${m}_{44}$	& 5.003 \\
%${m}_{45}$	& 4.529 \\
%${m}_{46}$	& 5.171 \\
%${m}_{47}$	& 4.852 \\
%${m}_{48}$  & 5.13 \\

\begin{figure}[]
    \centering
    \includegraphics[width=\linewidth]{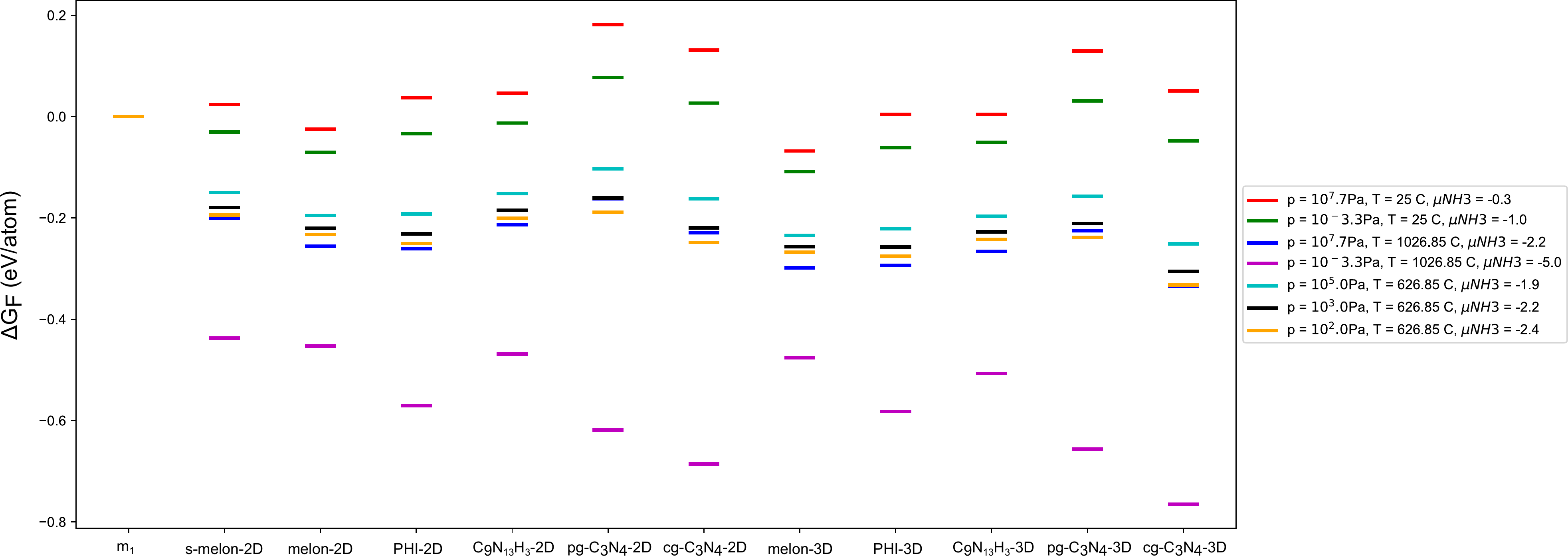}
    \caption{Free energy of formation for 2D and 3D structures with regards to various $\mu_\text{\ce{NH3}}$. The $\mu_\text{\ce{NH3}}$ is differentiated from the combination of $T$ and $p_{\ce{NH3}}$.  The free energy of formation is normalized to eV/atom due to the different numbers of elements.}
    \label{formation_free_overall_si}
\end{figure}

\begin{figure}[]
    \centering
    \includegraphics[width=\linewidth]{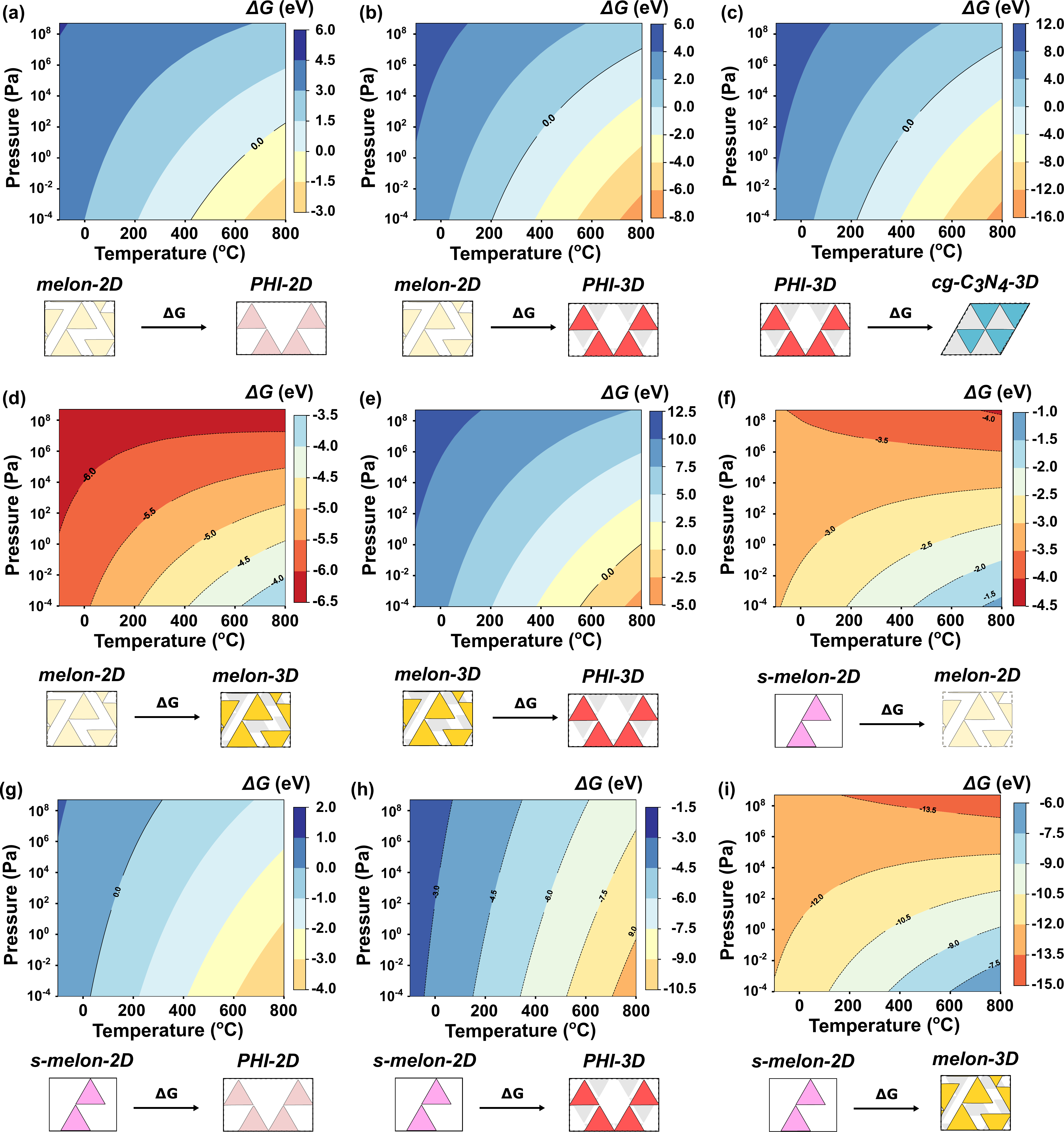}
    \caption{Respective phase diagram of the free energy reaction for 2D and 3D structures. The contour map is used to find spontaneous $\Delta\mu_\text{\ce{\ce{NH3}}}$ for temperature and $p_{\ce{NH3}}$ conditions.}
    \label{reaction_free_si}
\end{figure}

\begin{table}[]
\caption{Summary of Optical property}
\centering
\begin{tabular}{|c| c || c| c|}
 \hline
 \textbf{Structure} & \textbf{H-L gap (eV)} & \textbf{Structure} &  \textbf{bandgap (eV)}  \\
 \hline
 h$_{1}$	& 4.62 & s-melon-2D         & 3.62 \\
 h$_{2}$	& 3.98 & melon-2D       	& 3.68 \\
 h$_{31}$	& 3.64 & PHI-2D	            & 3.32 \\
 h$_{32}$	& 3.56 & \ce{C9M13H3}-2D	& 2.95 \\
 h$_{33}$	& 3.92 & cg-\ce{C3N4}-2D	& 2.99 \\
 h$_{34}$	& 3.80 & pg-\ce{C3N4}-2D	& 2.79 \\
 h$_{41}$	& 3.52 & s-melon-3D         & 3.39 \\
 h$_{42}$	& 3.71 & melon-3D	        & 3.12 \\
 h$_{43}$	& 3.52 & PHI-3D	            & 3.18 \\
 h$_{44}$	& 3.61 & \ce{C9M13H3}-3D    & 2.86 \\
 h$_{45}$	& 3.46 & cg-\ce{C3N4}-3D    & 2.95 \\
 h$_{46}$	& 3.65 & pg-\ce{C3N4}-3D    & 2.35 \\
 h$_{47}$	& 3.59 & Mixed1-3D          & 2.95 \\
 h$_{48}$   & 3.72 & Mixed2-3D          & 2.88 \\
            &      & Mixed3-3D          & 2.99 \\
            &      & Mixed4-3D          & 2.89 \\
 \hline
\end{tabular}
\label{table:table_optical_gap}
\end{table}

\begin{figure}[]
    \centering
    \includegraphics[width=\linewidth/3*2]{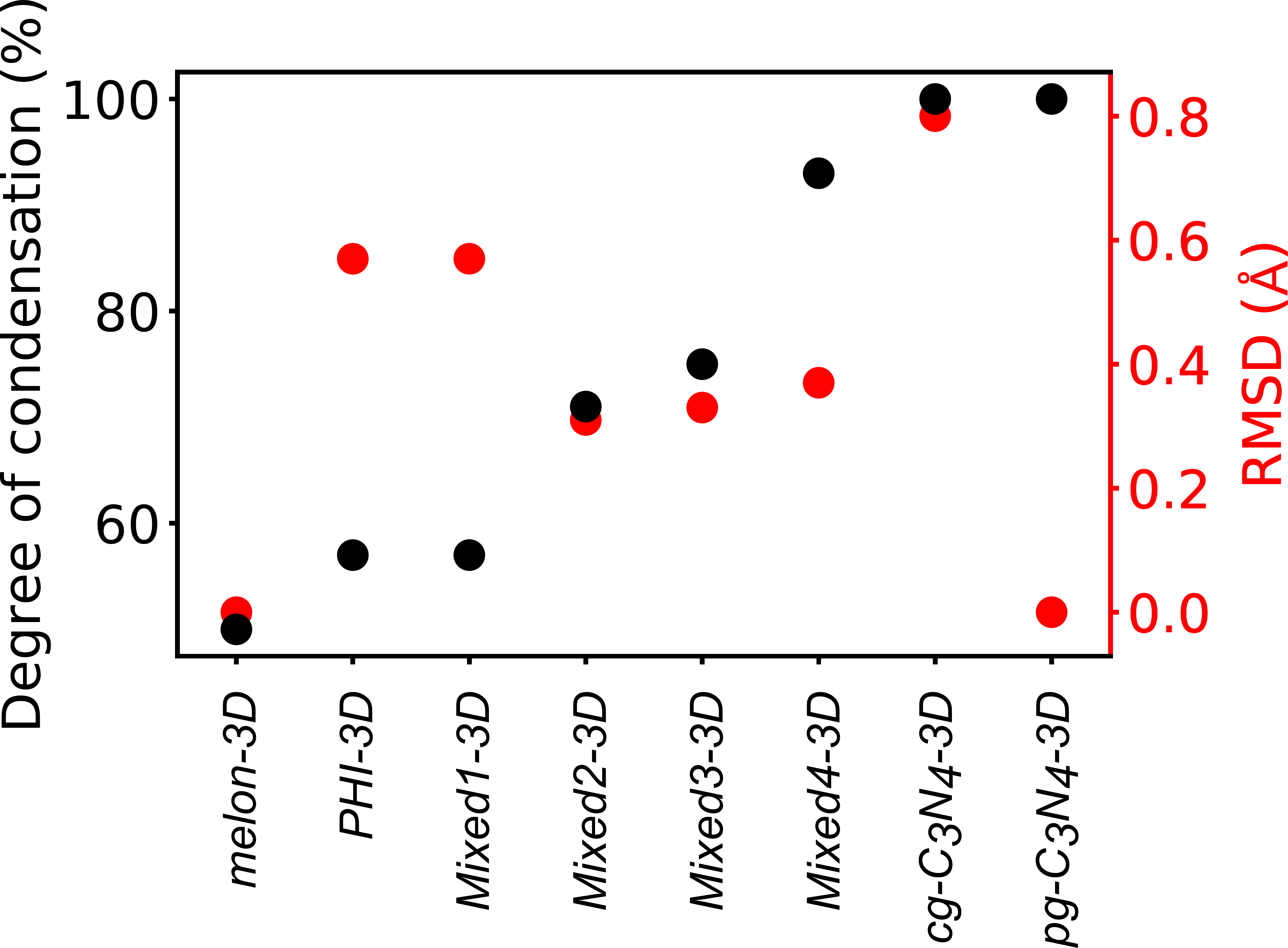}
    \caption{The degree of condensation and RMSD value are plotted with 3D structures.The Mixed structures are composed of PHI and g-\ce{C3N4} structures (see Figure \ref{list_of_mixed_structures}). The RMSD value of 3D structure is obtained from the average of the each layer.}
    \label{RMSD}
\end{figure}

\begin{table}[]
\caption{Summary of Degree of Corrugation}
\centering
\begin{tabular}{|c | c || c |c|}
 \hline
  \textbf{\multirow{2}{4em}{Structure}}&\textbf{RMSD of}&\textbf{\multirow{2}{4em}{Structure}}& \textbf{RMSD of}  \\
  &\textbf{out-of-plane (\AA)}& & \textbf{out-of-plane (\AA)}  \\
 \hline
 h$_{1}$	& 0.01 & melon-3D           & 0.00 \\
 h$_{2}$	& 0.08 & PHI-3D	            & 0.33 \\
 h$_{31}$	& 0.20 & cg-\ce{C3N4}-3D	& 0.80 \\
 h$_{41}$	& 0.20 & pg-\ce{C3N4}-3D	& 0.00 \\
 h$_{51}$	& 0.25 & Mixed1-3D          &  0.57 \\
 Mixed1-2D  & 0.58 & Mixed2-3D          &  0.57 \\
 Mixed3-2D 	& 0.35 & Mixed3-3D          &  0.31 \\
 PHI-2D     & 0.57 & Mixed4-3D          &  0.37 \\
 cg-\ce{C3N4}-2D & 0.81 & & \\
 \hline
\end{tabular}
\label{table:table_corrugation}
\end{table}

\end{suppinfo}
\section{Reference}
\bibliography{achemso}
\end{document}